\documentclass{article}
\usepackage{amssymb}
\usepackage{amsmath}
\usepackage{amsfonts}
\usepackage{geometry}
\usepackage{graphicx}
\usepackage{theorem}

\newtheorem{theorem}{Theorem}[section]

{\theorembodyfont{\normalfont}
\newtheorem{algorithm}[theorem]{Algorithm}

}

{\theorembodyfont{\normalfont}

}

{\theorembodyfont{\normalfont}

}
\newtheorem{definition}[theorem]{Definition}
{\theorembodyfont{\normalfont}

}

{\theorembodyfont{\normalfont}

}

\newtheorem{proposition}[theorem]{Proposition}
{\theorembodyfont{\normalfont}
\newtheorem{remark}[theorem]{Remark}

}

\begin{document}

\title{An Elementary, First Principles Approach to The Indefinite Spin Groups}

\author{E. Herzig \\
Department of Mathematical Sciences\\
University of Texas at Dallas\\
Richardson, TX 75080 \and V. Ramakrishna \\
Department of Mathematical Sciences\\
University of Texas at Dallas\\
Richardson, TX 75080\\
Corresponding Author} 
\date{}
\maketitle

\begin{abstract}
In this work we provide an elementary derivation of the indefinite
spin groups ${\mbox Spin}^{+}(p,q)$ in low-dimensions. Our approach
relies on the isomorphism ${\mbox Cl}(p+1, q+1) = M(2, Cl(p,q))$, simple
properties of Kronecker products, characterization of when an even
dimensional real (resp. complex) matrix represents a complex 
(resp. quaternionic) linear transformation, and basic aspects of
the isomorphism $\mathbf{H}\otimes \mathbf{H} = M(4, \mathbf{R})$. 
Of these the last is arguably the
most vital. Among other things it yields a surprisingly ubiquitous
role for the matrix $\left (\begin{array}{cc}
0 & i\sigma_{y}\\
i\sigma_{y} & 0
\end{array} \right )$. Our approach has the benefit of identifying these
spin groups as explicit groups of matrices within the same collection
of matrices which define $Cl(p, q)$. In other words, we do not work in
the algebra of matrices that the collection of even vectors  
in $Cl(p,q)$ is isomorphic to. 
This is crucial in some applications and also has the advantage
of being didactically simple. 
\end{abstract}

\section{Introduction}

The spin groups play a vital role in theoretical and applied aspects of
Clifford Algebras. Though the most popular spin group coverings are those
of the rotation groups, the ``indefinite"
coverings $SL(2, \mathbf{C}) \rightarrow SO^{+} (3,1, \mathbf{R}), 
Sp (4, \mathbf{R})\rightarrow 
SO^{+} (3,2, \mathbf{R})$
and $SU(2,2) \rightarrow SO^{+}(4, 2, \mathbf{R})$ are also important in 
many physical applications, \cite{goongi,rau,gilmore,pertii}.

In the literature the usual derivations of these covering groups usually
proceeds by working within the matrix algebra that the collection 
of even vectors in $Cl(p, q)$ is isomorphic to (see, for instance,
the excellent text \cite{portei}). Though elegant and important
such a derivation is inadequate for some applications
and often requires ad hoc constructions.  

One application that requires that the spin group be identified
as a collection of matrices within the same matrix algebra that the
one-vectors of $Cl(p,q)$ live in (thus the matrix algebra $Cl(p,q)$
is itself isomorphic to) is that of computing exponentials of matrices
in $so (p, q, \mathbf{R})$. Usage of the covering group provides an algorithm
(summarized in Algorithm 1 in the appendix ) which renders 
the problem of exponentiating such matrices tractable,
but which requires that the covering group lives 
in the same matrix algebra as the
one-vectors.

Didactically too, it is illuminating to be able to identify the 
spin group as an explicit group of matrices in the matrix algebra
$Cl(p, q)$ is isomorphic to. In other words, it is desirable to be
able to explicitly identify (not merely upto isomorphism) the spin group
as a collection of matrices in $Cl(p, q)$ emanating from the defining conditions
for the spin group [See IV) of Definition (\ref{basics}) ].

Related to this issue is the fact that the explicit form of the spin group
is very much dependent on the choice of one-vectors chosen for $Cl(p,q)$
(and hence on the attendant form of Clifford conjugation, 
reversion and grade morphisms).
Thus a statement such as ``${\mbox Spin}^{+}(3,2)$ is isomorphic
to $Sp(4, \mathbf{R})$" is more useful if one knows with respect
to which set of one-vectors
in $Cl(3, 2)$ (the double ring of $M(4, \mathbf{R})$) this statement pertains
to, and thus also how precisely $Sp(4, \mathbf{R})$ sits 
inside the subcollection
of $8\times 8$ real matrices that $Cl(3,2)$ is isomorphic to.

As a second illustration of this matter, consider the well known
fact that $SL(2, \mathbf{C})$ is the double cover of the Lorenz group. 
This is often demonstrated by first identifying $\mathbf{R}^{4}$
with $H_{2}$, the space of $2\times 2$ 
Hermitian matrices and then  observing that i) the quadratic
form associating to the generic element,
$tI_{2} + x \sigma{x} + y\sigma_{y} + z\sigma_{z}$ of $H_{2}$ minus
its determinant,
viz.,  $-{\mbox det}(tI_{2} + x \sigma{x} + y\sigma_{y} + z\sigma_{z})
= x^{2} + y^{z} + z^{2} - t^{2}$, coincides with the Minkowski metric;
and ii) the conjugation action of $SL(2, \mathbf{C})$ 
on  $H_{2}$ preserves this
quadratic form. 

Though undoubtedly
elegant and extremely useful in its own right, this derivation begs the  
question of how precisely $SL(2, \mathbf{C})$ arises from 
its Clifford theoretic
description.  

In this work we provide satisfactory resolutions to these questions.
The key ingredients in our approach are the following:

\begin{itemize}
\item Iterative use of the explicit isomorphism between $Cl(p+1, q+1)$ and
$M(2, Cl(p,q))$. This isomorphism provides a ready supply of a basis of
$1$-vectors for the various Clifford algebras considered here, starting
typically with the Pauli matrices (the quintessential anticommuting
collection of matrices).
The price (reward?) for using these bases of $1$-vectors is that the
associated version of the spin group is 
typically not the standard one. In order
to reconcile these versions with more familiar versions, the subsequent
tools mentioned below play an important role.

\item Characterization of when a $2n\times 2n$ real (resp. complex)
matrix represents a $n\times n$ complex (resp. quaternionic) linear
transformation. This characterization suggests how to modify
the basis of $1$-vectors provided by 
the first ingredient above, so as to render the
versions of the spin group closer to more familiar realizations.
More specifically,
this characterization suggests what the desirable form of the
grade automorphism ought to be in some cases, and the basis of 
$1$-vectors provided by the
first ingredient above is then modified accordingly.

\item Basic aspects of the Kronecker products of matrices.
The most pertinent use of this is that many of the calculations
required in arriving at the descriptions of the spin groups, can be reduced
to mere \underline{inspection} by making use of Kronecker products.

\item The most basic aspects of the very useful isomorphism $\mathbf{H}\otimes
\mathbf{H} 
= M(4, \mathbf{R})$. 
This is, in our view, the important most concomitant. In particular, it
provides a basis for $M(4, \mathbf{R})$ consisting of real orthogonal matrices
which are either symmetic or antisymmetric. It is the antisymmetric ones
which enable seeking alternatives to the standard representative of the
symplectic inner product on $\mathbf{R}^{4}$. Furthermore, this basis contains
the matrix  $\left (\begin{array}{cc}
0 & i\sigma_{y}\\
i\sigma_{y} & 0
\end{array} \right )$, which intervenes
in the description of many of the
spin groups in this work.
\end{itemize}

Thus, besides containing an elementary approach to the spin groups, this work
also provides novel and sometimes quaint representations of some the
classical groups which arise as the spin groups. In addition, it provides
useful interpretations for some of the matrices
in the $\mathbf{H}\otimes \mathbf{H}$ basis for
$M(4, \mathbf{R})$. In particular, the matrix $M_{1\otimes k}$ intervenes
in an essential and entirely \underline{natural} fashion 
in the description of several of
the spin groups. In this sense, this work is another contribution to
the list of applications of quaternions and Clifford algebras to algorithmic/
computational
linear algebra, \cite{cliffminpolyi,haconi,nii,
ni,niii,expistruc,expisufour,noncompactportion,minpolyi,jgspi}.

Special attention to ${\mbox Spin}^{+}(3,2)$ and ${\mbox Spin}^{+}(4,2)$,
is paid in this work to illustrate the general ideas. These two cases  
provide the most vivid instances of 
the ubiquitous role of the matrix
$\left (\begin{array}{cc}
0 & i\sigma_{y}\\
i\sigma_{y} & 0
\end{array} \right )$. 
For the former, we also
derive the explicit form of the Lie algebra isomorphism between
${\mbox spin}^{+}(3,2)$ and $so(3,2, \mathbf{R})$. The explicit form of this 
isomorphism is needed for computing exponentials, for instance. 
For brevity, 
derivations for the Lie algebra isomoprhism between 
other ${\mbox spin}^{+}(p,q)$ and $so(p,q, \mathbf{R})$ have been omitted.

The balance of this manuscript is organized as follows.
In the next section all preliminary material required
for this paper is collected. We call special attention to
Remarks (\ref{NotinEven}), 
(\ref{DiscernTraceFree}) and (\ref{GradeChangeBasis}).
The third section details the
construction of ${\mbox Spin}^{+}(3,2)$. The fourth section contains 
the construction of ${\mbox Spin}^{+}(4, 2)$. The fifth section develops
the remaining spin groups considered in this work - specifically
those for which $(p, q)\in \{ (2,1), (2,2), (3,1),
(4,1), (1, 5), (3,3)\}$. The next section offers
some conclusions. There is an appendix, which
records the algorithm
adapted from our earlier work, \cite{jgspi},
for the computation of matrix exponentials
in $so(p, q, \mathbf{R})$ 
using the explicit forms of the covering maps developed
here.

\section{Notation and Preliminary Observations}

\subsection{Notation}

We use the following notation throughout

\begin{description}
\item[1] $\mathbf{H}$ is the set of quaternions; $\mathbf{P}$ the
set of purely imaginary quaternions. Let $K$ be an associative algebra. Then 
$M(n,K)$ is just the set of $n\times n$ matrices with entries in $K$. 
For $ X\in M(n, K)$ with $K= \mathbf{C}$ or $\mathbf{H}$,
$X^{\ast }$ is the matrix obtained by
performing entrywise complex (resp. quaternionic) conjugation first, and
then transposition. For $K= \mathbf{C}$,
$\bar{X}$ is the matrix obtained by performing entrywise complex
conjugation.

\item[2] $J_{2n}=\left( 
\begin{array}{cc}
0_{n} & I_{n} \\ 
-I_{n} & 0_{n}%
\end{array}%
\right) $. Associated to $J_{2n}$ are i) $Sp (2n, K) =
\{ X\in M(2n, K): X^{T}J_{2n}X = J_{2n}\}$. Here $K$ is $R$ or $C$.
$Sp(2n, K)$ is a Lie group; ii) $sp(2n, K) =
\{ X\in M(2n, K): X^{T}J_{2n} = -J_{2n}X\}$.
$sp (2n, K)$ is the Lie algebra of $Sp(2n, K)$. 

\item[3] $\widetilde{J}_{2n}=J_{2}\oplus J_{2}\oplus \ldots \oplus J_{2}$.
Thus $\widetilde{J}_{2n}$ is the $n$-fold direct sum of $J_{2}$. $\widetilde{%
J}_{2n}$, is of course, explicitly permutation similar to $J_{2n}$, but it
is important for our purposes to maintain the distinction. Accordingly $%
\widetilde{Sp}(2n, K) =\{X\in M(2n, K):
\widetilde{J}_{2n}^{-1}X^{T}\widetilde{J}%
_{2n}=\widetilde{J}_{2n}\}$. 
Similarly, $\widetilde{sp}(2n, K) =\{X\in M(2n, K):
X^{T}\widetilde{J}_{2n}=-\widetilde{J}_{2n}X\}$.
Other variants of $J_{4}$ are of importance to this
paper, and they will be introduced later at appropriate points (see Remark %
\ref{The3Js} below).

\item[4] Let $p> 0, q > 0$ and $n= p+ q$.
Then  $I_{p,q} = {\mbox diag} (I_{p} - I_{q})$. Associated to
$I_{p,q}$ is the Lie algebra $so(p, q, R) = \{ X\in M(n, R):
X^{T}I_{p,q} = -I_{p, q}X\}$, sometimes called the Lorenz Lie algebra.
We defer definitions of the associated pertinent Lie groups to 
Section 2. 
The Lie group $SU(p, q)
= \{ X\in M(n, C): X^{*}I_{p,q}X = I_{p, q}, {\mbox det}(X) = 1\}$. 
Its Lie algebra is $su(p, q) = 
\{ X\in M(n, C): X^{*}I_{p,q} = -I_{p, q}X, {\mbox tr}(X) = 0\}$.

\item[5] The Pauli Matrices are 
\begin{equation*}
\sigma _{x}= \sigma_{1} =\left( 
\begin{array}{cc}
0 & 1 \\ 
1 & 0%
\end{array}%
\right) ;\text{ }\sigma _{y}=\sigma _{2}=\left( 
\begin{array}{cc}
0 & -i \\ 
i & 0%
\end{array}%
\right); \text{ }\sigma _{z}=\sigma _{3}=\left( 
\begin{array}{cc}
1 & 0 \\ 
0 & -1%
\end{array}%
\right) 
\end{equation*}

Note $i\sigma_{y} = J_{2}$.

\item[6] The matrix $K_{2l}$ is 
\begin{equation*}
K_{2l}=\left( 
\begin{array}{cc}
0_{l} & I_{l} \\ 
I_{l} & 0_{l}%
\end{array}%
\right) 
\end{equation*}%
This matrix enables succinct expressions for Clifford conjugation in
some of the Clifford algebras used in this work.
\end{description}

\subsection{Reversion, Clifford Conjugation and the Spin Groups}

The reader is referred to the excellent texts, 
\cite%
{pertii,portei} for formal definitions of Clifford algebras $Cl(p, q)$.
For our purposes it suffices to record the following: 

\begin{definition}
\label{basics}

{\rm \begin{description}
\item[I)] The reversion anti-automorphism on a Clifford algebra, $\phi ^{rev}
$, is the linear map defined by requiring that i) $\phi ^{rev}(ab)=\phi
^{rev}(b)\phi ^{rev}(a)$; ii) $\phi ^{rev}(v)=v$, for all $1$-vectors $v$;
and iii) $\phi ^{rev}(1)=1$. For brevity we will write $X^{rev}$ stands for
$\phi ^{rev}(X)$.

\item[II)] The Clifford conjugation anti-automorphism on a Clifford algebra, 
$\phi ^{cc}$, is the linear map defined by requiring that i) $\phi
^{cc}(ab)=\phi ^{cc}(b)\phi ^{cc}(a)$; ii) $\phi ^{cc}(v)=-v$, for all $1$%
-vectors $v$; and iii) $\phi ^{cc}(1)=1$. Once again $\phi ^{cc}(X)$ is also 
denoted as  $X^{cc}$.

\item[III)] The grade automorphism on a Clifford algebra, $\phi ^{gr}$ is $%
\phi ^{rev}\circ \phi ^{cc}$. 
Once again we write $X^{gr}$ for $\phi
^{gr}(X)$.

\item[IV)] $\mbox Spin^{+}( p, q) $ is the collection of elements $x$ in 
$Cl (p, q)$ satisfying the following
requirements: $i)$ $x^{gr}=x$, i.e., $x$ is even; $ii)$ $xx^{cc}=1$; and $%
iii)$ For all $1$-vectors $v$ in $Cl(p, q)$, $%
xvx^{cc}$ is also a $1$-vector. The last condition, in the presence of the
first two conditions, is known to be superfluous for $p + q\leq 5$, \cite%
{pertii,portei}. ${\mbox Spin}^{+}(p, q)$ is a connected Lie group, whose
Lie algebra is descibed in the next item.

\item[V)] ${\mbox spin}^{+}(p,q)$ is the Lie algebra of
$\mbox Spin^{+}( p, q) $. It is a fact that it also equals the space of
bivectors in $Cl (p, q)$, \cite{pertii,portei}. Thus if $\{X_{i}, i=1, 
\ldots , n\}$ is a basis of $1$-vectors for $Cl(p, q)$, then the space
of bivectors is the real span of $\{X_{k}X_{l}, k < l\}$. This space
does not, of course, depend on the choice of basis of $1$-vectors.
\end{description}
}
\end{definition}

\begin{remark}
\label{NotinEven}
{\rm The even subalgebra of $Cl(p, q)$ is $Cl(p, q-1)$ [respectively,
$Cl(q,p-1)$] if $q\geq 1$(respectively, if $p\geq 1$). Since the even
subalgebras are also Clifford algebras they too are matrix algebras.
However, in this work we prefer to view the even subalgebra
as a matrix subalgebra of the ambient matrix algebra $Cl(p, q)$. The reason
for this is that the analysis of the remaining conditions definining
${\mbox Spin}^{+}(p, q)$ will then \underline{not}
require any ad hoc constructions.}
\end{remark}

\begin{remark}
\label{DiscernTraceFree}
{\rm  Each $Cl(p, q)$ is a matrix algebra with entries in some associative
algebra suitably concocted out of the real numbers, the complex numbers or
the quaternions, \cite{pertii,portei}.  On this matrix algebra
the explicit matrix forms of reversion, grade and Clifford conjugation
depend in an essential fashion
on the choice of basis of $1$-vectors for $Cl(p, q)$. 
Once these forms have been obtained, the first
two conditions in the definition of ${\mbox Spin}^{+}(p, q)$ [IV) of Definition
(\ref{basics})] will lead 
to a group of matrices $\mathbf{K}$ in $Cl(p, q)$.
The third condition,
pertinent only when $p+q \geq 6$, imposes further restrictions
on this group. These additional restrictions
are best analysed by passing to the Lie algebra ${\mbox spin}^{+}(p, q)$.
In this work these restictions are analysed in one of two ways: I)
Suppose the first two conditions definining ${\mbox Spin}^{+}(p, q)$
lead to a representation $\Psi (L_{1})\subseteq Cl(p, q)$ of a Lie
algebra $L_{1}$ for the Lie algebra of $\mathbf{K}$. 
Then dimension
considerations often suggest $\Psi (L_{2})$, as a candidate
for ${\mbox spin}^{+}(p, q)$,
where $L_{2}$ is a codimension one Lie subalgebra of $L_{1}$. 
One then formally confirms
that $\Psi (L_{2})$ is indeed ${\mbox spin}^{+}(p, q)$ by showing that if $X
\neq 0$ is
in $\Psi (L_{1})$ 
but $X$ does not belong to $\Psi (L_{2})$, 
then $X$ violates the linearization of the third
condition in the definition of ${\mbox Spin}^{+}(p,q)$, i.e., one shows that
there is a one vector $v$ such that $Xv - vX$ is not a one-vector.
Typically $X$ will be the $\Psi$ image of a multiple of the identity 
matrix. 
II) Alternatively one computes explicitly the space of bivectors and attempts
to discern a recognizable structure of a Lie algebra of matrices on it.}
\end{remark}

\begin{remark}
\label{GradeChangeBasis}
{\rm The even subalgebra of $Cl(p, q)$ is evidently important in 
the description of the spin group. Therefore, to the extent posible,
it is vital to be able
to identify it as a ``recognizable" matrix subalgebra of the matrix
algebra $Cl(p, q)$ is isomorphic to. Equivalently it is useful to work
with a convenient form for the grade isomoprhism. However, this depends
on the choice of basis of one-vectors. In this work, starting with a basis
of $1$-vectors, the grade isomorphism will always take the form
$X\rightarrow M^{-1}\psi (X) M$, where $M$ is some unitary matrix and
$\psi (X)$ is either $X$ or $\bar{X}$ (where $\bar{X}$ is entrywise
complex or quaternionic conjugation). Suppose now that it is preferable
for the grade isomoprhism to read as $X\rightarrow N^{-1}\psi (X) N$,
with $N$ a matrix \underline{explicitly} conjugate to $M$, via $N = P^{-1}MP$.
This is then achieved by changing the original basis of one-vectors
$X_{i}$ to a new basis of $1$-vectors $Y_{i} = P^{-1}X_{i}P$. Furthermore,
suppose Clifford conjugation with respect to the $\{X_{i}\}$ reads as
$X\rightarrow C^{-1}\phi (X) C$, where $\phi (X)$ is either $X^{T}$ or
$X^{*}$ (where $*$ is Hermitian complex or quaternionic
conjugation) and $C$ some unitary matrix, as will be 
the case throughout this work.
Then Clifford conjugation, with respect to the new basis $\{Y_{i}\}$, is given
by $X\rightarrow D^{-1}\phi (X)D$, with $D = P^{-1}CP$.
It is emphasized that our techniques render
the finding of such a conjugation $P$ fully constructive - eigencomputations
going beyond $2\times 2$ matrices are never invoked.}
\end{remark}

\begin{definition}
\label{Lorenzgroup}
{\rm The Lorenz group, $SO^{+}(p, q, \mathbf{R})$ 
is the connected component of the
identity of the group $SO(p, q, \mathbf{R}) = \{X: X^{T}I_{p,q}X = I_{p, q}.
{\mbox det}(X) = 1\}$. Its Lie algebra (as well as the Lie algebra
of $SO(p, q, \mathbf{R})$ is $so (p, q, \mathbf{R})$. It is known that
the map which assigns to any element $g\in {\mbox Spin}^{+}(p, q)$ the
matrix of the linear map $v\rightarrow gvg^{cc}$, with $v$ a $1$-vector
in $Cl(p, q)$ is a $2:1$ group homomorphism from ${\mbox Spin}^{+}(p, q)$
to $SO^{+}(p, q, \mathbf{R})$. The map obtained by linearizing this map,
viz., the matrix of
$v\rightarrow Xv - vX$ (where $v$ is a $1$-vector and $X$ an element
of ${\mbox spin}^{+}(p,q)$), is a Lie algebra isomorphism
from ${\mbox spin}^{+}(p,q)$ to $so (p, q, \mathbf{R})$.}
\end{definition}
 
\subsection{An Iterative Construction}

Here we will outline an iterative construction of $1$-vectors for certain
Clifford Algebras, given a choice of one vectors for another Clifford
Algebra, \cite{pertii,portei}. Throughput this work this construction,
together with the contents of Remarks (\ref{ccandrevononemore}) below, is
collectively referred to as {\bf IC}.  

\begin{description}
\item[\textbf{IC}] ${\mbox C}l\left( p+1,\text{ }q+1\right) $ as $M\left( 2,%
\text{ }{\mbox C}l\left( p,\text{ }q\right) \right) $, where $M(2,\mathfrak{A%
})$ stands for the set of $2\times 2$ matrices with entries in an
associative algebra $\mathfrak{A}$: Suppose $\{e_{1},\ldots
,e_{p},f_{1},\ldots ,f_{q}\}$ is a basis of $1$-vectors for ${\mbox C}%
l\left( p,\text{ }q\right) $. So, in particular, $e_{k}^{2}=+1,$ $k=1,\ldots
,p$ and $f_{l}^{2}=-1,$ $l=1,\ldots ,q$. Then a basis of $1$-vectors for ${%
\mbox C}l\left( p+1,\text{ }q+1\right) $ is given by the following
collection of elements in $M(2,{\mbox C}l\left( p,\text{ }q\right) )$: 
\begin{equation*}
\left( 
\begin{array}{cc}
e_{k} & 0 \\ 
0 & -e_{k}%
\end{array}%
\right) ,\text{ }k=1,\ldots ,p;\text{ }\left( 
\begin{array}{cc}
0 & 1 \\ 
1 & 0%
\end{array}%
\right) ;\text{ }\left( 
\begin{array}{cc}
f_{l} & 0 \\ 
0 & -f_{l}%
\end{array}%
\right) ,\text{ }l=1,\ldots ,q;\text{ }\left( 
\begin{array}{cc}
0 & 1 \\ 
-1 & 0%
\end{array}%
\right) 
\end{equation*}%
The $1$ and the $0$ in the matrices above are the identity and zero elements
of ${\mbox C}l\left( p,\text{ }q\right) $ respectively.

\end{description}

\begin{remark}
\label{ccandrevononemore}If Clifford conjugation and reversion have been
identified on ${\mbox C}l\left( p,\text{ }q\right) $ with respect to some
basis of $1$-vectors, then there are explicit expressions for Clifford
conjugation and reversion on ${\mbox C}l\left( p+1,\text{ }q+1\right) $ with
respect to the basis of $1$-vectors described in iterative construction 
\textbf{IC1} above. Specifically if $X=\left( 
\begin{array}{cc}
A & B \\ 
C & D%
\end{array}%
\right) $, then we have 
\begin{equation*}
X^{CC}=\left( 
\begin{array}{cc}
D^{rev} & -B^{rev} \\ 
-C^{rev} & A^{rev}%
\end{array}%
\right) 
\end{equation*}%
while reversion is 
\begin{equation*}
X^{rev}=\left( 
\begin{array}{cc}
D^{cc} & B^{cc} \\ 
C^{cc} & A^{cc}%
\end{array}%
\right) 
\end{equation*}%
This is immediate from the definitions of reversion and Clifford conjugation.

Note that if elements of ${\mbox C}l\left( p,\text{ }%
q\right) $ have been identified with $l\times l$ matrices, then 
\begin{equation*}
X^{cc}=J_{2l}^{-1}[\left( 
\begin{array}{cc}
A^{rev} & B^{rev} \\ 
C^{rev} & D^{rev}%
\end{array}%
\right) ]^{BT}J_{2l}
\end{equation*}%
and that 
\begin{equation*}
X^{rev}=K_{2l}^{-1}[\left( 
\begin{array}{cc}
A^{cc} & B^{cc} \\ 
C^{cc} & D^{cc}%
\end{array}%
\right) ]^{BT}K_{2l}
\end{equation*}%
where $K_{2l}$ is the matrix at the end of Section $2.1$, and if $X=\left( 
\begin{array}{cc}
Y & Z \\ 
U & V%
\end{array}%
\right) $ is a $2\times 2$ block matrix, then $X^{BT}=\left( 
\begin{array}{cc}
Y & U \\ 
Z & V%
\end{array}%
\right) $
\end{remark}

\subsection{$\protect\theta _{%
\mathbb{C}
}$ and $\protect\theta _{\mathbb{H}}$ matrices:}

Some of the material here is to be found in \cite{hhorni}, for instance.

\begin{definition}
\label{complexmatrices}Given a matrix $M\in M\left( n,\text{ }%
\mathbb{C}
\right) $, define a matrix $\theta _{%
\mathbb{C, I}
}(M)\in M\left( 2n,\text{ }%
\mathbb{R}
\right) $ by first setting $\theta _{%
\mathbb{C, I}
}(z)=\left( 
\begin{array}{cc}
x & y \\ 
-y & x%
\end{array}%
\right) $ for a complex scalar $z=x+iy$. We then define $\theta _{%
\mathbb{C}
}(M)=(\theta _{%
\mathbb{C}
}(m_{ij}))$, i.e., $\theta _{%
\mathbb{C}
}(M)$ is a $n\times n$ block matrix, with the $(i,j)$th block equal to the $%
2\times 2$ real matrix $\theta _{%
\mathbb{C}
}(m_{ij})$.
\end{definition}

\begin{remark}
\label{Cproperties}Properties of $\theta _{%
\mathbb{C, I}
}$:  Some useful useful properties of the map $\theta _{{\bf C, I}}$
now follow:

\begin{description}
\item[i)] $\theta _{{\bf C, I}}$
is an $\mathbf{R}$-
linear map.

\item[ii)] $\theta _{{\bf C, I}}$
is an associative algebra isomorphism onto its image from
$M(n, \mathbf{C})$ to $M(2n, \mathbf{R})$.

\item[iii)] $\theta _{{\bf C, I}}
(M^{\ast })=[\theta _{{\bf C, I}}
(M)]^{T}$

\item[iv)] $\theta _{{\bf C, I}}
(I_{n})=I_{2n}$

\item[v)] A useful property is the following: $X\in M( 2n, \mathbf{R})$
is in the image of $\theta _{\bf{C, I}}$
iff $X^{T}=\widetilde{J}_{2n}^{-1}X^{T}\widetilde{J}_{2n}$
iff $X\widetilde{J}_{2n} = X\widetilde{J}_{2n}$.
\end{description}
\end{remark}

\begin{remark}
{\rm We call an $X\in {\mbox Im}(\theta_{{\bf C, I}}$
a $\theta _{{\bf C, I}}$
matrix.} 

\end{remark}

\begin{remark}
\label{ThetaIIC}
{\rm An closely related  alternative way to asociate real matrices to
matrices in $M(n, \mathbf{C})$ is as follows. Write $X\in M(n, \mathbf{C})$
as $X = Y + iZ$ with $Y, Z$ both real matrices. The define
$\Theta_{\bf{C, II}}(Z) = \left ( \begin{array}{cc}
X & - Y\\
Y & Z
\end{array}
\right )$. All contents of Remark (\ref{Cproperties}) apply verbatim
except v) which is now replaced by ``$X\in M(2n, \mathbf{R})$ is in the
image $\Theta_{\bf{C, II}}$ iff $XJ_{2n} = J_{2n}X$.}
\end{remark}

\noindent Next, to a matrix with quaternion entries will be associated a
complex matrix. First, if $q\in \mathbb{H}$ is a quaternion, it can be
written uniquely in the form $q=z+wj$, for some $z,$ $w\in 
\mathbb{C}
$. Note that $j\eta =\bar{\eta}j$, for any $\eta \in 
\mathbb{C}
$. With this at hand, the following construction associating complex
matrices to matrices with quaternionic entries (see \cite{hhorni} for
instance) is useful:

\begin{definition}
Let $X\in M(n,\mathbb{H})$. By writing each entry $x_{pq}$ of $X$ as%
\begin{equation*}
x_{pq}=z_{pq}+w_{pq}j,\text{ }z_{pq},w_{pq}\in 
\mathbb{C}
\end{equation*}%
we can write $X$ uniquely as $X=Z+Wj$ with $Z,$ $W\in M\left( n,\text{ }%
\mathbb{C}
\right) $. Associate to $X$ the following matrix $\theta _{\mathbb{H}}(X)\in
M\left( 2n,\text{ }%
\mathbb{C}
\right) $: 
\begin{equation*}
\theta _{\mathbb{H}}(X)=\left( 
\begin{array}{cc}
Z & W \\ 
-\bar{W} & \bar{Z}%
\end{array}%
\right) 
\end{equation*}
\end{definition}

\noindent Next some useful properties of the map $\theta _{\mathbb{H}}:M(n,%
\mathbb{H})\rightarrow M\left( 2n,\text{ }%
\mathbb{C}
\right) $ are collected.

\begin{remark}
\label{Hproperties} Properties of $\theta _{\mathbb{H}}$:

\begin{description}
\item[i)] $\theta _{\mathbb{H}}$ is an $\mathbf{R}$
linear map.

\item[ii)] $\theta _{\mathbb{H}}(XY)=\theta _{\mathbb{H}}(X)\theta _{\mathbb{%
H}}(Y)$

\item[iii)] $\theta _{\mathbb{H}}(X^{\ast })=[\theta _{\mathbb{H}}(X)]^{\ast
}$. Here the $\ast $ on the left is quaternionic Hermitian conjugation,
while that on the right is complex Hermitian conjugation.

\item[iv)] $\theta _{\mathbb{H}}(I_{n})=I_{2n}$

\item[v)] A less known property is the following: $\Lambda \in M(2n,
\mathbf{C} )$
is in the image of $\theta _{\mathbb{H}}$ iff $\Lambda ^{\ast
}=J_{2n}^{-1}X^{T}J_{2n}$.
\end{description}
\end{remark}

\begin{remark}
{\rm We call an $\Lambda \in {\mbox Im}(\theta _{\mathbf{H}})$,
a $\theta_{\mathbf{H}}$ matrix. 
In \cite{hhorni} such matrices are called matrices of
the quaternion type. But we eschew this nomenclature.} 
\end{remark}

\begin{definition}
\label{SLNH}
{\rm The determinant of a matrix $X\in M(n, \mathbf{H})$ is defined
to be the determinant of the complex matrix $\Theta_{\mathbf{H}}(X)$.
$SL(n, \mathbf{H})$ is the group of $n\times n$ quaternionic matrices
with unit determinant. Its Lie algebra, $sl(n, \mathbf{H})$ is the
subset of $M(n, \mathbf{H})$ satisfying ${\mbox Re} ({\mbox Tr} \ X) =
0$.}
\end{definition}

\begin{remark}
\label{CommutewithDiagonal}
{\rm Though the contents of this remark seem unrelated to the rest
of this subsection, they play a similar utilitarian role in this work.
Specifically, a matrix with entries in ${\mathbf C}$ commutes with a diagonal
matrix whose equal entries appear contiguously, if and only if the
matrix is block-diagonal, \cite{hhorni}. In this work this diagonal
matrix will be $I_{r, s}$ (in which case this result applies even
for matrices with entries in $\mathbf{H}$).
This observation will be used, when applicable, to change the
basis of $1$-vectors via permutation similarities so as to render the even
subalgebra equal to an algebra of block-diagonal matrices.}
\end{remark}

\subsection{$\mathbf{H}\otimes \mathbf{H}$ and $M(4, \mathbf{R})$} 

 The algebra isomorphism between between $\mathbb{H}\otimes \mathbb{H}$ and $%
M\left( 4,\text{ }%
\mathbb{R}
\right) $ (also denoted by $gl(4,%
\mathbb{R}
)$) may be summarized as follows:

\begin{itemize}
\item Associate to each product tensor $p\otimes q\in \mathbb{H}\otimes 
\mathbb{H}$, the matrix, $M_{p\otimes q}$, of the map which sends $x\in 
\mathbb{H}$ to $px\bar{q}$, identifying $%
\mathbb{R}
^{4}$ with $\mathbb{H}$ via the basis $\left\{ 1,\text{ }i,\text{ }j,\text{ }%
k\right\} $. Here, $\bar{q}=q_{0}-q_{1}i-q_{2}j-q_{3}k$

\item Extend this to the full tensor product by linearity. This yields an
associative algebra isomorphism between $\mathbb{H}\otimes \mathbb{H}$ and $%
M\left( 4,\text{ }%
\mathbb{R}
\right) $. Furthermore, a basis for $gl(4,%
\mathbb{R}
)$ is provided by the sixteen matrices $M_{e_{x}\otimes e_{y}}$ as $e_{x},$ $%
e_{y}$ run through $1,$ $i,$ $j,$ $k$.

\item We define conjugation on $\mathbb{H}\otimes \mathbb{H}$ by setting $%
\bar{p\otimes q}=\bar{p}\otimes \bar{q}$ and then extending by linearity.
Conjugation in $\mathbb{H}\otimes \mathbb{H}$ corresponds to matrix
transposition, i.e., $M_{\bar{p}\otimes \bar{q}}=(M_{p\otimes q})^{T}$. A
consequence of this is that any matrix of the form $M_{1\otimes p}$ or $%
M_{q\otimes 1}$, with $p,$ $q\in \mathbb{P}$ is a real antisymmetric matrix.
Similarly, the most general special orthogonal matrix in $M\left( 4,\text{ }%
\mathbb{R}
\right) $ admits an expression of the form $M_{p\otimes q}$, with $p$ and $q$
both unit quaternions.
\end{itemize}

\begin{remark}
\label{The3Js}
{\rm Some important matrices from this basis for $M(4, \mathbf{R})$
provided by $\mathbb{H}\otimes \mathbb{H}$ are: 

\begin{itemize}
\item $M_{1\otimes j}$ is precisely $J_{4}$.

\item The matrix $M_{1\otimes k}$, which we denote by $\breve{J}_{4}$.
Note also that
$M_{1\otimes k}$ is a Kronecker product, viz.,
$\sigma_{x}\otimes i\sigma_{y}$.

\item $M_{i\otimes 1} = -\widetilde{J}_{4}$.
\end{itemize} }

\end{remark}

\subsection{Kronecker Products}

The Kronecker product $A\otimes B$, \cite{hhorni} has the following
properties which will be used throughout this work. 
\begin{itemize}
\item {\bf KP1} $(A\otimes B)(C\otimes D) = AC \otimes BD$. $(A\otimes B)^{T} =
A^{T}\otimes B^{T}$.

\item {\bf KP2} A special case of {\bf KP1}, worth recording separately, is
the following. If $(\lambda , v)$ and $(\mu, w)$ are eigenpairs of 
$A_{n\times n}$ and $B_{m\times m}$ respectively, 
then $(\lambda\mu, v\otimes w)$ is an eigenpair
for $A\otimes B$. This observation will be used in Section 4 to produce
a certain desired conjugation by mere inspection. 
\item {\bf KP3} If $A$ and $B$ are square then ${\mbox Tr}(A\otimes B)
= {\mbox Tr}(A)
{\mbox Tr}(B)$.
\end{itemize}

\section{${\mbox Spin}^{+}(3,2)$}

It is convenient to begin with $\{\sigma_{z}\}$ as  a basis of $1$-vectors for 
${\mbox C}l\left( 1,\text{ }0\right) $. 
Quite clearly reversion with respect to this basis
is $X\rightarrow X^{T}$. Similarly, Clifford conjugation is
$X\rightarrow J_{2}^{T}X^{T}J_{2}$. Applying {\bf IC} to this basis
yields the following basis for ${\mbox C}l\left ( 2,\text{ }1\right )$:
\begin{equation}
\label{Basis21}
\{ \sigma_{z}\otimes \sigma_{z}, \sigma_{x}\otimes I_{2},
i\sigma_{y}\otimes I_{2}\}
\end{equation} 

Hence by {\bf IC} 
reversion on ${\mbox C}l\left( 2,\text{ }1\right) $,
with respect to this basis is 
\begin{equation}
\label{Rev21}
X\rightarrow R_{2,1}^{T}X^{T}R_{2,1};
R_{2,1} = \left (\begin{array}{cc}
0_{2} & J_{2}\\
J_{2} & 0_{2}
\end{array} \right ) = M_{1\otimes k}
\end{equation}

Clifford conjugation on ${\mbox C}l\left( 2,\text{ }1\right) $, 
with respect to this basis, is 
\begin{equation}
\label{Cliff21} 
X\rightarrow C_{2,1}^{T}
X^{T}C_{2,1}, 
C_{2,1} = J_{4}
\end{equation}

 Applying {\bf IC} again produces the following basis of $1$-vectors
for ${\mbox C}l\left ( 3,\text{ }2\right )$:
\begin{equation}
\label{Basis32}
\{ \sigma_{z}\otimes\sigma_{z}\otimes\sigma_{z}, 
\sigma_{z}\otimes\sigma_{x}\otimes I_{2},
\sigma_{x}\otimes I_{2}\otimes I_{2},
\sigma_{z}\otimes i\sigma_{y}\otimes I_{2},
i\sigma_{y}\otimes I_{2}\otimes I_{z}\}
\end{equation}

As is well known ${\mbox C}l\left( 3,\text{ }2\right) $ is the
double ring of $M(4, \mathbf{R})$. Specifically, the last basis
renders ${\mbox C}l\left( 3,\text{ }2\right) $ equal to the subalgebra
of $M(8, \mathbf{R})$, consisting of matrices which when written
as $4\times 4$ block matrices, have each block equal to a \underline{diagonal}
$2\times 2$ matrix.

Reversion with respect to this last basis for 
${\mbox C}l\left( 3,\text{ }2\right)$ is thus given by

\begin{equation}
\label{Rev32}
X\rightarrow R_{3,2}^{T}X^{T}R_{3,2}; R_{3,2} = \left (\begin{array}
{cc} 0_{4} & J_{4}\\
J_{4} & 0_{4}
\end{array}
\right )
\end{equation}

Clifford conjugation on ${\mbox C}l\left( 3,\text{ }2\right)$ is
provided  
is expressible as
\begin{equation}
\label{Cliff32}
X\rightarrow C_{3,2}^{T}X^{T}C_{3,2};
C_{3,2} = \left ( \begin{array}{cc}
0_{4} & M_{1\otimes k}\\
-M_{1\otimes k} & 0_{4}
\end{array}
\right )
\end{equation}
  
Hence the grade involution on ${\mbox C}l\left( 3,\text{ }2\right)$ is
given by
\begin{equation}
\label{Grade32}
X \rightarrow G_{3,2}^{T}XG_{3,2}, G_{3,2} =
\left ( \begin{array}{cc}
-M_{1\otimes i} & 0_{4}\\
0_{4} & M_{1\otimes i}
\end{array}
\right )
\end{equation}

Hence ${\mbox Spin}_{+}(3,2)$ equals the subalgebra of 
matrices $X\in M(8, \mathbf{R})$ characterized by the following conditions:
\begin{itemize}
\item $X = (X_{ij}), i, j=1, \ldots , 4$, with each
$X_{ij}$ a $2\times 2$ diagonal matrix.
\item $G_{3,2}$ commutes with $X$.
\item $X^{T}C_{3,2}X = C_{3,2}$
\end{itemize}

The first two conditions, as a calculation shows,
impose the following restrictions on the blocks
of $X$:
\begin{itemize}
\item $X_{11}, X_{14}, X_{22}, X_{23}, X_{32}, X_{33}, X_{41}$ and
$X_{44}$ are $2\times 2$ skew-Hamiltonian matrices, and thus $2\times 2$
constant multiples of $I_{2}$. 
\item The remaining $X_{ij}$ are $2\times 2$, diagonal, Hamiltonian
matrices. Since $2\times 2$ Hamiltonian matrices are precisely the
traceless $2\times 2$ matrices, this forces each of these $X_{ij}$ to
be a constant multiple of $\sigma_{z}$.
\end{itemize}

Hence $X$ being an even element of 
${\mbox C}l\left( 3,\text{ }2\right)$ is equivalent to asserting that
it has the following form
\[
X = \left (\begin{array}{cccccccc}
a & 0 & b & 0 & c & 0 & d & 0\\
0 & a & 0 & -b & 0 & -c & 0 & d\\
e & 0 & f & 0 & g & 0 & h & 0\\
0 & -e & 0 & f & 0 & g & 0 & -h\\
i & 0 & j & 0 & k & 0 & l & 0\\
0 & -i & 0 & j & 0 & k & 0 & -l\\
m & 0 & n & 0 & p & 0 & q & 0\\
0 & m & 0 & -n & 0 & -p & 0 & q
\end{array}
\right )
\]
 
Let us now examine the condition $C_{3,2}^{T}X^{T}C_{3,2}X
= I_{8}$.
Some calculations which  systematically use the Hamiltonian/skew-Hamiltonian
structures of the blocks
$X_{ij}$ of $X$ reveals the following conditions to be equivalent to
$C_{3,2}^{T}X^{T}C_{3,2}X
= I_{8}$:

\begin{enumerate}
\item The conditions corresponding to the equality of the $(1,1)$ blocks and
also the $(4,4)$ blocks are identical to 
$aq-dm + ih - el = 1$.
\item The conditions corresponding to the equality of the $(1,2)$ blocks and
also the $(3,4)$ blocks are identical to      
to $qb - fl + hj - dn = 0$.

\item The equality of the $(1,3)$ blocks and also the $(2,4)$
blocks yield the condition 
$kh-dp + qc - lg = 0$.

\item The equality of the $(2,1)$ blocks and also the $(4,3)$ blocks
yield the condition 
$cm - gi + ke -ap = 0$.

\item The equality of the $(2,2)$ blocks and similarly that of
the $(3,3)$ blocks impose the condition 
$cn - gj + kf -bp = 1$.
 
\item The equality of the $(3,1)$ blocks and similarly that of 
the $(4,2)$ blocks impose the condition 
$fi - bm +  an - je = 0$.
   
\item The equality of the remaining blocks holds automatically and
thus imposes no further restriction on the entries of $X$. 
\end{enumerate}

Let us extract information from these conditions as follows.
Define a $4\times 4$ matrix associated to an even vector $X$ as follows:

\begin{equation}
\label{TheZEquation}
Z =  \left ( \begin{array}{cccc}
a & b & c & d\\
e & f & g & h\\
i & j & k & l\\
m & n & p & q
\end{array}
\right )
\end{equation}

Note the relation between $Z$ and $X$ can be rewritten via a map
$\Lambda$ as
\begin{equation}
\label{TheLambda}
X = \Lambda (Z) =  
\left ( \begin{array}{cccc}
aI_{2} & b\sigma_{z} & c\sigma_{z} & dI_{2}\\
e\sigma_{z} & fI_{2} & gI_{2} & h\sigma_{z}\\
i\sigma_{z} & jI_{2} & kI_{2} & l\sigma_{z}\\
mI_{2} & n\sigma_{z} & p\sigma_{z} & qI_{2}
\end{array}
\right )
\end{equation}

$\Lambda$ is an algebra isomorphism of
$M(4, \mathbf{R})$ onto its image in $M(8, \mathbf{R})$ which furthermore
satisfies $\Lambda (Z^{T}) = [\Lambda (Z)]^{T}$.

Let us also collect the six distinct relations characterizing
the $X^{cc}X = I$ condition again for ease of reference:
\begin{eqnarray}
\label{SixConditions1}
aq - dm + ih - el & = & 1\\ \nonumber
bq - fl + hj -dn & = & 0\\ \nonumber
kh - dp + qc - gl & = & 0\\ \nonumber
cm - gi + ke - ap & = & 0\\ \nonumber
cn - gj + kf - bp & = & 1\\ \nonumber
fi - bm + an - je & = & 0
\end{eqnarray}

Then, after some experimentation and invocation of the
the basis of $M(4, \mathbf{R})$ provided by its isomorphism
with $\mathbf{H}\otimes \mathbf{H}$, the conditions
in Equation (\ref{SixConditions1}) are seen to be equivalent to 
$Z$ belonging to a \underline{nonstandard}
representation of $Sp(4, \mathbf{R})$. Specifically,
defining \[
\breve{J}_{4} = M_{1\otimes k} = \left (\begin{array}{cc}
0_{2} & J_{2}\\
J_{2} & 0_{2}
\end{array}
\right )
\]
the conditions in Equation (\ref{SixConditions1}) are equivalent
to
\begin{equation}
Z^{T}\breve{J}_{4}Z = \breve{J}_{4}
\end{equation}

To see this it is better to write 
$Z$ in Equation (\ref{TheZEquation}) as a $2\times 2$
block matrix, viz., 
$Z = \left (\begin{array}{cc}
A & B\\
C & D
\end{array}
\right )$ [so that $A  = \left (\begin{array}{cc}
a & b\\
e & f 
\end{array}
\right )$; $B = \left (\begin{array}{cc}
c & d\\
g &h 
\end{array}
\right )$; $C = \left (\begin{array}{cc}
i & j \\
m & n
\end{array}
\right )$; and $D =  
 \left (\begin{array}{cc}
k & l\\
p & q 
\end{array}
\right )$]. Then the condition $Z^{T}\breve{J}_{4}Z = \breve{J}_{4}$
is equivalent to:
\begin{itemize}
\item $C^{T}J_{2}A + A^{T}J_{2}C = 0$, i..e., $C^{T}J_{2}A$ is symmetric. 
\item $C^{T}J_{2}B + A^{T}J_{2}D = J_{2}$.
\item $D^{T}J_{2}A + B^{T}J_{2}C = J_{2}$. Note this condition is 
identical to the previous one, since $J_{2}$ is antisymmetric.
\item $D^{T}J_{2}B + B^{T}J_{2}D = O_{2}$, i.e., $D^{T}J_{2}B$ is symmetric.   
\end{itemize}

Now the $(1,2)$ entry and the $(2,1)$ entry of $C^{T}J_{2}A$ are 
$-bm + if$ and $-an + je$ respectively. Thus $C^{T}J_{2}A$ being symmetric
is precisely the sixth of the conditions in Equation (\ref{SixConditions1}).
Next, the $(1,2)$ entry and the $(2,1)$ entry of $D^{T}J_{2}B$ are
$-pd + kh$ and $-qc + lg$ respectively. Thus $D^{T}J_{2}B$ being symmetric
is precisely the third of the conditions in Equation (\ref{SixConditions1}).      
  
Finally, a direct calculation shows that 
$C^{T}J_{2}B + A^{T}J_{2}D = J_{2}$ is equivalent to the first, second,
fourth and the fifth conditions in 
in Equation (\ref{SixConditions1}).   
 
Thus, indeed ${\mbox Spin}^{+}(3, 2)$, with respect to the
basis of $1$-vectors in Equation (\ref{Basis32}),
is isomorphic to a nonstandard representation of $Sp(4, R)$.
Thus we are lead to the following
\begin{theorem}
\label{ComprehensiveSpin}
{\rm Consider the basis of $1$-vectors for $Cl(3, 2)$, given
by $\{X_{1}, \ldots , X_{5}\} = \{ 
\sigma_{z}\otimes\sigma_{z}\otimes\sigma_{z}, 
\sigma_{z}\otimes\sigma_{x}\otimes I_{2},
\sigma_{x}\otimes I_{2}\otimes I_{2},
\sigma_{z}\otimes i\sigma_{y}\otimes I_{2},
i\sigma_{y}\otimes I_{2}\otimes I_{z}\}$. With respect to this basis,
\begin{itemize}
\item i) Clifford Conjugation is given by Equation (\ref{Cliff32}). 
\item ii) Reversion is given by Equation (\ref{Rev32}).
\item iii) Grade involution is given by Equation (\ref{Grade32}). 
\item ${\mbox Spin}^{+}(3, 2)$ is given by the subset of
$M(8, R)$, admitting a representation of the form
\[
\Lambda (X) = 
\left (\begin{array}{cccc}
aI_{2} & b\sigma_{z} & c\sigma_{z} & dI_{2}\\
e\sigma_{z} & fI_{2} & gI_{2} & h\sigma_{z}\\
i\sigma_{z} & jI_{2} & kI_{2} & l\sigma_{z}\\
mI_{2} & n\sigma_{z} & p\sigma_{z} & qI_{2}
\end{array}
\right )
\]

where the real $4\times 4$ matrix  
\[
X = \left (\begin{array}{cccc}
a & b & c & d\\
e & f & g & h\\
i & j & k & l\\  
m & n & p & q
\end{array}
\right )
\]
satisfies $X^{T}M_{1\otimes k} X = M_{1\otimes k}$. The collection of
such $4\times 4$ matrices, denoted $\breve{Sp}(4, \mathbf{R})$, is isomorphic
to (the standard representation of) $Sp(4, \mathbf{R})$, via an explicit
conjugation.
\end{itemize}
}
\end{theorem}

Let us now use this to compute explicitly the isomorphism between
$\breve{sp}(4, \mathbf{R}) = \{ Y\in M(4, \mathbf{R}): Y^{T}M_{1\otimes k}
= - M_{1\otimes k}Y\}$ and $so (3, 2, \mathbf{R})$. To this end
we proceed in 3 steps:
\begin{itemize}
\item Use as the basis of $\breve{sp}(4, R)$ the collection of matrices
$Z_{i}, i = 1, \ldots , 10$, where each $Z_{i}$ equals $M_{1\otimes k}W_{i}$,
with $W_{i}, i=1, \ldots, 10$ ranging over the set $\{I_{4}\}
\cup \{M_{a\otimes b}\}$, where $a, b$ range over the set $\{i, j, k\}
\subseteq H$. Note this latter union is a basis of all the symmetric
real $4\times 4$ matrix. 

\item Next embed each $Z_{i} =
\left (\begin{array}{cccc}
a & b & c & d\\
e & f & g & h\\
i & j & k & l\\  
m & n & p & q
\end{array}
\right )
$ as the $8\times 8$ matrix 
\begin{equation}
\Lambda (Z_{i})
= 
\left (\begin{array}{cccc}
aI_{2} & b\sigma_{z} & c\sigma_{z} & dI_{2}\\
e\sigma_{z} & fI_{2} & gI_{2} & h\sigma_{z}\\
i\sigma_{z} & jI_{2} & kI_{2} & l\sigma_{z}\\
mI_{2} & n\sigma_{z} & p\sigma_{z} & qI_{2}
\end{array}
\right )
\end{equation}

\item Then compute the $5\times 5$ matrices $M_{i}, i=1, \ldots , 10$,
of the linear maps, $X\rightarrow \theta(Z_{i})X - X\theta (Z_{i})$
on the space of $1$-vectors for $Cl (3, 2)$, with respect to the basis
$\{X_{1}, \ldots , X_{5}\} = \{ 
\sigma_{z}\otimes\sigma_{z}\otimes\sigma_{z}, 
\sigma_{z}\otimes\sigma_{x}\otimes I_{2},
\sigma_{x}\otimes I_{2}\otimes I_{2},
\sigma_{z}\otimes i\sigma_{y}\otimes I_{2},
i\sigma_{y}\otimes I_{2}\otimes I_{z}\}$ for the $1$-vectors of $Cl(3,2)$.
We note here that this computation is greatly facilitated by the fact that
each of the $\theta (Z_{i})$ also admits an expression as a triple
Kronecker product of (essentially) the Pauli matrices. 
{\it This again attests to the remarkable utility of the $H\otimes H$
basis for $M(4, \mathbf{R})$}.
\end{itemize} 

This leads to the following theorem

\begin{theorem}
\label{LieIsoTable}
{\rm The Lie algebra isomorphism $ \Psi_{3,2}: \breve{sp}(4, \mathbf{R})
\rightarrow 
so(3, 2, \mathbf{R})$ is given by the following table, wherein the second
column forms a basis for $\breve{sp}(4, \mathbf{R})$ and the last column 
forms a basis
for $so(3, 2, {R})$:

\begin{tabular}{cccc}
$W_{i}$ & $Z_{i}$ & $\Lambda (Z_{i})$ & $M_{i}$\\
$I_{4}$ & $M_{1\otimes k}$ & $\sigma_{x}\otimes (i\sigma_{y})\otimes I_{2}$ &
$2(e_{5}e_{4}^{T} - e_{4}e_{5}^{T})$\\
$M_{i\otimes i}$ & $M_{i\otimes j}$ & $-i\sigma_{y}\otimes i\sigma_{y}
\otimes I_{2}$
& $-2(e_{4}e_{3}^{T} + e_{3}e_{4}^{T})$\\
$M_{j\otimes j}$ & $-M_{j\otimes i}$ & $\sigma_{x}\otimes \sigma_{x}
\otimes I_{2}$ &
$2(e_{5}e_{2}^{T} + e_{2}e_{5}^{T})$\\
$M_{k\otimes k}$ & $-M_{k\otimes 1}$ & $i\sigma_{y}\otimes \sigma_{x}
\otimes I_{2}$ &
$2(e_{2}e_{3}^{T} - e_{3}e_{2}^{T})$\\
$M_{i\otimes j}$ & $-M_{i\otimes i}$ & $-\sigma_{z}\otimes I_{2}\otimes 
I_{2}$ &
$-2(e_{5}e_{3}^{T} + e_{3}e_{5}^{T})$\\
$M_{i\otimes k}$ & $-M_{i\otimes 1}$ & $I_{2}\otimes i\sigma_{y}
\otimes\sigma_{z}$ &
$2(e_{1}e_{2}^{T} - e_{2}e_{1}^{T})$\\
$M_{j\otimes i}$ & $-M_{j\otimes j}$ & $I_{2}\otimes \sigma_{z}
\otimes I_{2}$ &
$2(e_{4}e_{2}^{T} - e_{2}e_{4}^{T})$\\
$M_{j\otimes k}$ & $-M_{j\otimes 1}$ & $i\sigma_{y}\otimes \sigma_{z}\otimes
\sigma_{z}$ & $2(e_{1}e_{3}^{T} - e_{3}e_{1}^{T})$\\
$M_{k\otimes i}$ & $M_{k\otimes j}$ & $I_{2}\otimes \sigma_{x}\otimes 
\sigma_{z}$
& $-2(e_{4}e_{1}^{T} + e_{4}e_{1}^{T})$\\
$M_{k\otimes j}$ & $-M_{k\otimes i}$ & $\sigma_{x}\otimes\sigma_{z}\otimes
\sigma_{z}$
& $2(e_{5}e_{1}^{T} - e_{1}e_{5}^{T})$ 
\end{tabular}

}
\end{theorem} $\diamondsuit$

{\bf Proof:}
The proof is a detailed calculation following the strategy outlined before
the statement of the theorem. We will provide the details for the first
row of the table:
\noindent  $W_{1} = I_{4}$, $Z_{1} = M_{1\otimes k}$,
$\Lambda (Z_{1}) = \sigma_{x}\otimes (i\sigma_{y})\otimes I_{2}$.
Then $\Lambda (Z_{1})X_{i} - X_{i}\Lambda (Z_{1}) = 0$ for all
$i\in \{1,2,3,4,5\}$ except $i= 4,5$ when we, in fact, have
$\Lambda (Z_{1})X_{4} - X_{4}\Lambda (Z_{1}) = 2X_{5}$ and
$\Lambda  (Z_{1})X_{5} - X_{5}\Lambda  (Z_{1}) = -2X_{4}$.
Thus $M_{1} = 2 (e_{5}e_{4}^{T} - e_{4}e_{5}^{T})$.
$\diamondsuit$

\begin{remark}
\label{exp32}
Exponentiating Matrices in $so (3, 2, \mathbf{R})$: {\rm Theorems 
 (\ref{ComprehensiveSpin}) and (\ref{LieIsoTable}), together with
the Algorithm in the appendix provide a constructive procedure,
requiring no eigencomputations, for exponentiating a matrix
$X\in so (3, 2, \mathbf{R})$. We first compute the element
$Z = \Psi_{3,2}^{-1}(X)\in \breve{sp}(4, \mathbf{R})$ using the
II and IV column of the table in Theorem (\ref{LieIsoTable}).
Suppose the matrix $e^{Z}\in \breve{Sp}(4, \mathbf{R})$ has been computed.
Then the matrix $\Lambda (e^{Z})\in Cl (3, 2)\subseteq M(8, \mathbf{R})$
is found (this being a matter of mere inspection). Then one computes
\[
 M_{i} = 
[\Lambda (e^{Z})]X_{i}[C_{3,2}^{T}\Lambda (e^{Z^{T}})C_{3,2}],
i=1, \ldots ,5
\]
with $X_{i}$ as in Theorem (\ref{ComprehensiveSpin}) and $C_{3,2}$
as in Equation (\ref{Cliff32}). Note use has been made of the fact
that the embedding $\Lambda$ in Equation (\ref{TheLambda}) satisfies
$[\Lambda (X)]^{T} = \Lambda (X^{T})$. The computation of $M_{i}$ is 
straightforward in view
of the form of the $8\times 8$ matrix $\Theta (e^{Z})$. 
Then $M_{i} = \sum_{j=1}^{5}g_{ji}X_{i}$,
where the real numbers $g_{ji}$ can be computed by using the trace
inner product (this computation being facilitated by the fact that
the one vectors $X_{i}$ are triple Kronecker products). The matrix
$(g_{ij}), i, j =1, \ldots , 5$ then is \underline{precisely $e^{X}$}.
So everything reduces to the computation of $e^{Z}$ for $Z\in
\breve{sp}(4, \mathbf{R})$. But this can also be done explicitly
in closed form. Indeed, if $q$ is any unit quaternion such that
$\bar{q}kq = j$, then the matrix $M_{1\otimes q}^{T}ZM_{1\otimes q}$ 
is a $4\times 4$
Hamiltonian matrix and hence their minimal polynomials and exponentials
can be computed, \underline{without} eigencomputations 
in closed form. Hence so too can $e^{Z}$. Passage to
$sp (4, \mathbf{R})$ can be eschewed by directly working with $Z$
and using the methods of
of \cite{minpolyi} if one prefers.}    
\end{remark}  

\section{${\mbox Spin}^{+}(4, 2)$}

One begins with ${\mbox Cl}(2, 0)$, represented by the following
basis of $1$-vectors:
\begin{equation}
\label{Basis20}
\{ \sigma_{x}, \sigma_{z}\}
\end{equation}
In view of the fact that $\sigma_{x}, \sigma_{z}$ are symmetric it follows
that reversion on $Cl(2, 0)$ is
\begin{equation}
\label{Revon20}
X \rightarrow X^{T}
\end{equation} 
 
Similarly since each of $\sigma_{x}, \sigma_{z}$
is similar to minus itself via $J_{2} = i\sigma_{y}$ it follows that
that, with respect to this basis of $1$-vectors, Clifford conjugation
on ${\mbox Cl}(2, 0)$ is the map 
\begin{equation}
\label{Cliffon20}
X\rightarrow J_{2}^{T}X^{T}J_{2}
\end{equation}

Now we apply {\bf IC} to the foregoing
to produce the following basis of $1$-vectors
for ${\mbox Cl}(3,1)$
\begin{equation}
\label{FirstBasis31}
\{\sigma_{z}\otimes \sigma_{x}, \sigma_{z}\otimes\sigma_{z},
\sigma_{x}\otimes I_{2}, i\sigma_{y}\otimes I_{2}\}
\end{equation}.

{\bf IC} together with some block matrix calculations shows that reversion
on $Cl(3,1)$, with respect to the basis in Equation (\ref{FirstBasis31}),
is given by
\[
X\rightarrow P_{4}^{T}X^{T}P_{4}
\]
where $P_{4} = M_{1\otimes k}$ again.

Similarly,  Clifford conjugation on ${\mbox Cl}(3,1)$       
is $X\rightarrow J_{4}^{T}X^{T}J_{4}$.
Thus the grade involution on ${\mbox Cl}(3,1)$ is therefore
$X\rightarrow (-M_{1\otimes i})^{T}X(-M_{1\otimes i})$.

Let us now apply {\bf IC} again to the above basis of $1$-vectors
and forms of reversion and Clifford conjugation on
${\mbox Cl}(3,1)$ to produce the
following basis of $1$-vectors for ${\mbox Cl}(4,2)$:
\begin{equation}
\label{FirstBasis42}
\{ \sigma_{z}\otimes\sigma_{z}\otimes\sigma_{x},
\sigma_{z}\otimes\sigma_{z}\otimes\sigma_{z},
\sigma_{z}\otimes\sigma_{x}\otimes I_{2},
\sigma_{x}\otimes I_{2}\otimes I_{2} = K_{8},
\sigma_{z}\otimes i\sigma_{y}\otimes I_{2},
i\sigma_{y}\otimes I_{2} \otimes I_{2} = J_{8}\}
\end{equation} 

Then reversion on ${\mbox Cl}(4,2)$ is, with respect to this basis 
of $1$-vectors, given by

\begin{equation}
\label{FirstRev42}
X\rightarrow P_{8}^{T}X^{T}P_{8}
\end{equation}
where $P_{8}$ is the product $\left (\begin{array}{cc}
J_{4} & 0_{4}\\
0_{4} & J_{4}
\end{array}
\right )K_{8} =
\sigma_{x}\otimes i\sigma_{y}\otimes
I_{2}$.

Similarly, Clifford conjugation 
 ${\mbox Cl}(4,2)$ is, with respect to this basis
of $1$-vectors, given by

\begin{equation}
\label{FirstCliff42}
X\rightarrow M_{8}^{T}X^{T}M_{8}, 
M_{8} = i\sigma_{y}\otimes \sigma_{x}\otimes i\sigma_{y}
\end{equation}

Therefore, the grade involution ${\mbox Cl}(4,2)$ is, 
with respect to the basis in Equation (\ref{FirstBasis42}), 
\begin{equation}
\label{FirstGrade42}
X\rightarrow G_{8}^{T}XG_{8}, 
G_{8} = 
{\mbox diag} (-J_{2}, J_{2}, J_{2}, -J_{2})
\end{equation}
where, in arriving at the form $G_{8}$, systematic use of Equation
{\bf KP1} of Section 2.6  was made. 

In view of the map $\theta_{\mathbf{C, I}}$, we would like the grade involution
to be $X\rightarrow G_{4,2}^{T}XG_{4,2}$, with $G_{4,2}
= \widetilde{J}_{8}$.

Since 
$\sigma_{z}^{-1}J_{2}\sigma_{z}
= \sigma_{z}J_{2}\sigma_{z} = -J_{2}$,
it is seen that $G_{8}$ and $G_{4,2}$ are explicitly conjugate
\[
S^{-1}G_{8}S = G_{4,2}
\]
with
\[
S =  S^{-1} = S^{T} = {\mbox diag}[\sigma_{z}, I_{2}, I_{2}, \sigma_{z}]
\]

Accordingly we also change the basis of $1$ -vectors for
${\mbox Cl}(4, 2)$. If we let the basis members in Equation
(\ref{FirstBasis42}) by $\{X_{i}, i=1, \ldots , 6\}$,
then we define a second basis of $1$-vectors for ${\mbox Cl}(4, 2)$ by
$Y_{i} = S^{-1}X_{i}S, i=  1, \ldots , 6$.

This leads to the following basis:
\begin{equation}
\label{IIBasisfor42}
\{-\sigma_{z}\otimes \sigma_{z}\otimes \sigma_{x}, 
\sigma_{z}\otimes \sigma_{z}
\otimes \sigma_{z}, \sigma_{z}\otimes \sigma_{x}\otimes \sigma_{z},
\sigma_{x}\otimes I_{2}\otimes \sigma_{z}, \sigma_{z}\otimes i\sigma_{y}
\otimes \sigma_{z}, i\sigma_{y}\otimes \sigma_{z}\}
\end{equation}     

Then, in view of Remark (\ref{GradeChangeBasis}),
the grade involution on $Cl(4,2)$ is indeed given by conjugation with respect
to $G_{4,2}$ as desired.

Defining, $C_{4,2} = S^{T}M_{8}S$, we find by Remark
(\ref{GradeChangeBasis}) that Clifford conjugation is,
with respect to the basis in Equation (\ref{IIBasisfor42}),
given by $Y\rightarrow C_{4,2}^{T}Y^{T}C_{4,2}$,  $\forall Y\in
{\mbox Cl}(4,2)$.

Now 
\begin{equation}
\label{NewCliff42}
C_{4,2}
= 
\left (
\begin{array}{cccc}
0_{2} & 0_{2} & 0_{2} & -J_{2}\\
0_{2} & 0_{2} & J_{2} & 0_{2}\\
0_{2} & -J_{2} & 0_{2} & 0_{2}\\
J_{2} & 0_{2} & 0_{2} & 0_{2}
\end{array}
\right )
\end{equation}

From the definition of the map $\Theta_{{\bf C, I}}$ we see that
\begin{equation}
\label{BreveAgain}
C_{4,2}
=
\Theta_{\mathbf{C, I}}[\left (
\begin{array}{cccc}
0 & 0 & 0 & i\\
0 & 0 & -i & 0\\
0 & i & 0 & 0\\
-i & 0 & 0 & 0
\end{array}
\right )]
= \Theta_{\mathbf{C, I}}[iM_{1\otimes k}]
\end{equation}

\begin{remark}
\label{breveagain}
{\rm In other words, the matrix $\breve{J}_{4} = M_{1\otimes k}$ once again
{\it intervenes in the form of Clifford conjugation} for $Cl(4,2)$, as it
did for $Cl(3,2)$ - a fact that would be difficult to arrive at 
without the $\mathbf{H}\otimes \mathbf{H}$ basis for $M(4, \mathbf{R})$!}
\end{remark}

Since $p + q > 5$, the spin group of $Cl(4,2)$ 
is the subset of those elements in 
$\{ Y\in M(8, \mathbf{R}) , 
YG_{4,2} = G_{4,2}Y, C_{4,2}^{T} Y^{T}C_{4,2}Y = I_{8}\}$
which satisfy the further restriction that they leave the space
of $1$-vectors invariant. To identify that subset we work at the level of
the Lie algebras, cf., Remark (\ref{DiscernTraceFree}). 
Specifically, dimension considerations show that
the Lie algebra ${\mbox spin}^{+}(4,2)$ must be a $15$-dimensional
Lie subalgebra of  

\begin{equation}
\label{Prespin42}
\{ Y\in M(8, \mathbf{R}) , YG_{4,2} = G_{4,2}Y, C_{4,2}^{T} Y^{T}
= -YC_{4,2}\},
\end{equation}
 which leaves the space of $1$-vectors invariant
under the action $Z\rightarrow ZY - YZ$, where $Y$ belongs to the set
in Equation (\ref{Prespin42}), and $Z$ is a $1$-vector. Equivalently  
it suffices to find an element in the set in Equation (\ref{Prespin42})
which violates this condition. The obvious candidate for this element
is $\Theta_{C, I}(iI_{4})$. To that end let $\{Y_{1}, \ldots , Y_{6}\}$
be the basis in Equation (\ref{IIBasisfor42}) and check if even one  
${\mbox Tr}\{Y_{k}^{T}[\Theta_{C, I}(iI_{4})Y_{i}
- Y_{i}\Theta_{C, I}(iI_{4})]\}$,
for some fixed $i \in \{1, \ldots , 6\}$ and each $k =1, \ldots , 6$ is
zero. Using the fact that $\Theta_{C, I}(iI_{4})$ is the {\it triple
Kronecker product} $I_{2}\otimes I_{2}\otimes i\sigma_{y}$ and that
the matrices $Y_{i}$ are {\it themselves triple Kronecker products} 
it is seen,
in fact, that these traces vanish for all $i=1, \ldots , 6$. This shows that
the Lie algebra of the spin group must be the $\Theta_{C, I}$ images of
matrices in the following set:
\[
\{ Z\in M(4, \mathbf{C}):   (iM_{1\otimes k})^{*}Z^{*} = - Z(iM_{1\otimes k});
{\mbox Tr}(Z) = 0\}
\]

Let us next relate these calculations to $SU(2,2)$ being isomorphic to 
${\mbox Spin}^{+}(4,2)$.
To that end the main observation is that the Hermitian matrix,
$iM_{1\otimes k}$,
is unitarily conjugate to $I_{2,2}$. This unitary conjugation
can, in fact, be explicitly found and this relies on the fact 
relies on the fact that $M_{1\otimes k}$ also equals the {\it Kronecker
product},  
$\sigma_{x}\otimes (i\sigma_{y})$.
Explicit orthonormal eigenpairs for each of the $2\times 2$ matrices
$\sigma_{x}$ and $i\sigma_{y}$ are easily found. They are
\begin{itemize}
\item $\sigma_{x}$: $\{ (1, u_{1}), (-1, u_{2})\}$
where $u_{1} = 
\frac{1}{\sqrt{2}}\left (\begin{array}{c}
1\\
1
\end{array}
\right )$
and $u_{2} = 
\frac{1}{\sqrt{2}}\left (\begin{array}{c}
1\\
-1
\end{array}
\right )$.

\item $i\sigma_{y}$: $\{ (i, v_{1}), (-i, v_{2})\}$,
where $v_{1} = \frac{1}{\sqrt{2}}\left (\begin{array}{c}
1\\
i
\end{array}
\right )$ and
$v_{2} =
\frac{1}{\sqrt{2}}\left (\begin{array}{c}
1\\
-1
\end{array}
\right )$.
\end{itemize}

Therefore from elementary properties of Kronecker products, it follows
that an explicit orthonormal eigenpairs for $C_{4,2}$ are given by
\[
\{ (1, u_{1}\otimes v_{2}), (1, u_{2}\otimes v_{1}),
(-1, u_{1}\otimes v_{1}), (-1, u_{2}\otimes v_{2})\}
\]
Therefore $U^{*}C_{4,2}U = I_{2,2}$, where
\[
U = [u_{1}\otimes v_{2} \mid u_{2}\otimes v_{1} 
\mid u_{1}\otimes v_{1} \mid  u_{2}\otimes v_{2}]
\]

\begin{equation}
\label{KroneckerUnitaryConjugation} 
U = \frac{1}{2}
\left (\begin{array}{cccc}
1 & 1 & 1 & 1\\
-i & i & i & -i\\
1 & -1 & 1 & -1\\
-i & -i & i & i
\end{array}
\right )
\end{equation}

These calculations can now be summarized as

\begin{theorem}
{\rm $Cl(4,2) = M(8, R)$ has the matrices $Y_{i}$ given by Equation
(\ref{IIBasisfor42}) as a basis of $1$-vectors. With respect to this basis
\begin{itemize}
\item The grade involution is $Y\rightarrow \widetilde{J}^{T}_{8}Y
\widetilde{J}_{8}$.
\item Clifford conjugation is given by $Y\rightarrow 
[\Theta_{{\bf C,I}}(iM_{1\otimes k})]^{T}Y^{T}
[\Theta_{\bf {C,I}}(iM_{1\otimes k})]$.
\item ${\mbox Spin}^{+}(4, 2)$ equals those matrices in
$Y\in M(8, R)$ satisfying the conditions: i) $Y$ is a $\Theta_{{\bf C, I}}$
matrix; ii) ${\mbox det}(Y) = 1$; 
iii) $[\Theta_{\bf {C, I}}(iM_{1\otimes k})]^{T}Y^{T}
[\Theta_{{\bf C, I}}(iM_{1\otimes k})]Y
= I_{8}$.
\item ${\mbox spin}^{+}(4,2)$ consists of those matrices $Y\in M(8, R)$
satisfying the following conditions:
i) $Y$ is a $\Theta_{{\bf C,I}}(Z)$ matrix with $Z\in M(4, \mathbf{C})$   
matrix; ii) ${\mbox Tr}(Z) = 0$;          
iii) $[\Theta_{C, I}(iM_{1\otimes k})]^{T}Y^{T}
= -Y[\Theta_{C, I}(iM_{1\otimes k})]$.
\end{itemize}

The group of matrices in $M(4, \mathbf{C})$ whose 
$\Theta_{{\bf C, I}}$ images equal ${\mbox Spin}^{+}(4,2)$ 
is explicitly conjugate
to $SU(2,2)$ via the 
matrix $U$ in Equation (\ref{KroneckerUnitaryConjugation}).}
\end{theorem}

\section{Other Low Dimensional ${\mbox Spin}^{+}(p,q)$}

In this appendix our approach to ${\mbox Spin}^{+}(p,q)$, when 
$(p, q)\in \{ (2,1), (2,2), (3,1),
(4,1), (1, 5), (3,3)\}$ is outlined. Of particular note is the completely
\underline{straightforward} derivation of the covering 
$SL(2, \mathbf{C})\rightarrow
SO^{+}(3, 1, \mathbf{R})$ and the somewhat surprising concentric embedding of
$SL(2, \mathbf{R})\times SL(2, \mathbf{R})$ in 
$SL(4, \mathbf{R})$ as ${\mbox Spin}^{+}(2,2)$.   
   
\subsection{${\mbox Spin}^{+}(2,1)$}

A basis of $1$-vectors for $Cl (2,1)$ is given by Equation (\ref{Basis21}).
Quite clearly ${\mbox Cl}(2,1)$ is the algebra of $2\times 2$ block
matrices with each block a $2\times 2$ real diagonal matrix.

Reversion and Clifford conjugation with respect to this basis are given
by Equations (\ref{Rev21}) and (\ref{Cliff21}) respectively.
So, the grade involution is $G_{2,1}(X) = (J_{4}M_{1\otimes k})^{T}X
(J_{4}M_{1\otimes k}) = M_{1\otimes i}^{T}XM_{1\otimes i}$.

Hence, writing an $X\in Cl(2,1)$ in the form $X  = \left (\begin{array}{cc}
A & B\\
C & D 
\end{array}
\right )$, with each of $A, B, C, D$ $2\times 2$ real diagonal, it is seen
that
$X$ is even iff
iff $A$ and $D$ commute with $J_{2}$ and $B$ and $C$ anti-commute
with $J_{2}$. So $A =a\sigma_{z}; B = bI_{2}; C= cI_{2}; D =d\sigma_{z}$.
Such an even $X$ belongs to ${\mbox Spin}^{+}(2,1)$ iff it also belongs
to $Sp(4, \mathbf{R}) = \{X\in M(4, \mathbf{R}) : 
J_{4}^{T}X^{T}J_{4}X
= I_{4}\}$. But, this is equivalent to $B^{T}D$ and $A^{T}C$ symmetric
and $A^{T}D - C^{T}D = I_{2}$. Since $A, B, C, D$ are all diagonal
the first two conditions are superfluous. Similarly, the 
diagonality of $A, \ldots , D$, shows that 
$A^{T}D - C^{T}D = (ad-bc)I_{2}$. 
So we conclude that an even $X\in
{\mbox Cl}(2,1)$ is in   
${\mbox Spin}^{+}(2,1)$ iff $ad-bc = 1$.

Thus ${\mbox Spin}^{+}(2,1)$ is isomorphic to $SL(2, \mathbf{R})$. Specifically,
$Spin^{+}(2,1)$ is the group of $4\times 4$ real matrices obtained
by embedding 
$\left (\begin{array}{cc}
a & b\\
c & d
\end{array}
\right )\in SL(2, \mathbf{R})$, in the form 
\[
\left (\begin{array}{cccc}
a & 0 & b & 0\\
0 & -a & 0 & -b\\
c & 0 & d & 0\\
0 & c & 0 & -d
\end{array}
\right )
\] 

\subsection{${\mbox Spin}^{+}(3,1)$}
The most natural basis of $1$-vectors for ${\mbox Cl}(2,0)$
\[
\{\sigma_{z}, \sigma_{x}\}
\]
With respect to this basis reversion on ${\mbox Cl}(2,0)$ is $X\rightarrow
X^{T}$, while Clifford conjugation is $X\rightarrow J_{2}^{T}X^{T}J_{2}$.

Application of {\bf IC} then produces the following:
\begin{enumerate}
\item A basis of $1$-vectors for ${\mbox Cl}(3,1) = M(4, \mathbf{R})$ given by
\[
\{\sigma_{z}\otimes\sigma_{z},\sigma_{z}\otimes\sigma_{x},
\sigma_{x}\otimes I_{2},i\sigma_{y}\otimes I_{2}\}
\]
\item With respect to this last basis reversion on ${\mbox Cl}(3,1)$
is $X = \left (\begin{array}{cc}
A & B\\
C & D
\end{array}
\right ) \rightarrow K_{4}^{T}
 \left (\begin{array}{cc}
J_{2}^{T}A^{T}J_{2} & J_{2}^{T}C^{T}J_{2}\\
J_{2}^{T}B^{T}J_{2} & J_{2}^{T}D^{T}J_{2}
\end{array}
\right )K_{4}  
= M_{1\otimes k}^{T}X^{T}M_{1\otimes k}
$
\item Clifford conjugation is $X\rightarrow J_{4}^{T}X^{T}J_{4}$
\item The grade automorphism is therefore $X\rightarrow 
(-M_{1\otimes i})^{T}X(-M_{1\otimes i})$
\end{enumerate}

Since ${\mbox Cl}(3,1)$ is $M(4, \mathbf{R})$ we would like the
grade automorphism to be given by $J_{4}$ instead of
$-M_{1\otimes i}$. To that end, we need a unit quaternion
$q$ such that $\bar{q}(-i)q = j$. One choice is $q =
\frac{1}{\sqrt{2}} (1 + k)$. Now $J_{4} = M_{1\otimes q}^{T}
(-M_{1\otimes i}) M_{1\otimes q}$. 
Accordingly we change the basis of $1$-vectors for
${\mbox Cl}(3,1)$ to $\{Y_{i}\}$, where $Y_{i}
= M_{1\otimes q}^{T}X_{i}M_{1\otimes q}$, with the $X_{i}$ the basis
of $1$-vectors above. 

Then invoking Remark (\ref{GradeChangeBasis}), it is seen that:
\begin{enumerate}
\item  With respect to this basis the grade involution is precisely
$X\rightarrow J_{4}^{T}XJ_{4}$.
\item 
Clifford conjugation is, with respect to this basis of $1$-vectors 
precisely $X\rightarrow M_{1\otimes i}^{T}X^{T}M_{1\otimes i}$.
\end{enumerate}

Thus, 
\begin{equation}
{\mbox Spin}^{+} (3,1) =
\{ Y\in M(4, \mathbf{R}): YJ_{4} = J_{4}Y; 
M_{1\otimes i}^{T}Y^{T}M_{1\otimes i}Y = I_{4}\}
\end{equation}

Remark (\ref{ThetaIIC}) yields that the 
first condition is equivalent to $Y = \left (\begin{array}{cc} 
A & -B\\
B & A
\end{array}
\right )$, with $A, B\in M(2, \mathbf{R})$.

The second condition says
\[
\left (\begin{array}{cc}
-J_{2}A^{T}J_{2}A + J_{2}B^{T}J_{2}B & J_{2}A^{T}J_{2}B + J_{2}B^{T}J_{2}A\\
-J_{2}B^{T}J_{2}A - J_{2}A^{T}J_{2}B & J_{2}B^{T}J_{2}B - J_{2}A^{T}J_{2}A 
\end{array}
\right ) = I_{4}
\]

Let $A = \left (\begin{array}{cc}
a & b\\
c & d
\end{array}
\right )$ and $B = \left (\begin{array}{cc}
\alpha & \beta \\
\gamma & \delta
\end{array}
\right )$.

Then saying
$Y\in {\mbox Spin}^{+}(3,1)$ is equivalent to the following four conditions:

\begin{itemize}
\item $(1,1)$ block $=I_{2}$, i.e.,
$-J_{2}A^{T}J_{2}A + J_{2}B^{T}J_{2}B = I_{2}$. 
A quick calculation shows that this is equivalent to
$ad-bc + \beta\gamma -\alpha\delta = 1$.

\item $(1,2)$ block $=0_{2}$, i.e., $J_{2}A^{T}J_{2}B +
J_{2}B^{T}J_{2}A = 0_{2}$. 
Equivalently,  
$b\gamma -\alpha d + \beta c - a\delta = 0$ 

\item $(2,1)$ block $= 0_{2}$. But the $(2,1)$
block, $-J_{2}B^{T}J_{2}A - J_{2}A^{T}J_{2}B$ is precisely minus the
$(1,2)$ block. So this condition is now redundant.

\item $(2,2)$ block $= I_{2}$. But the $(2,2)$ block,  
$J_{2}B^{T}J_{2}B - J_{2}A^{T}J_{2}A$, is identical to the $(1,1)$
block.
\end{itemize}       

Now consider ${\mbox det}(A + iB) =
(a+i\alpha )(d+i\delta ) - (b + i\beta )(c +i\gamma )
= (ad-\alpha \delta + \beta\gamma - bc) + i(a\delta + \alpha d - b\gamma 
- \beta c) = = + i0 = 1$.
Thus, indeed ${\mbox Spin}^{+}(3,1)$ is the $\Theta_{{\bf C, II}}$ image of
$SL(2, \mathbf{C})$ in $M(4, \mathbf{R})$, and hence 
isomorphic as a group to
$SL(2, \mathbf{C})$.

\subsection{${\mbox Spin}^{+}(2,2)$}
Observing first that a  basis of $1$-vectors for $Cl (1,1)$ is
$\{\sigma_{x}, i\sigma_{y}\}$, with respect to which reversion and
Clifford conjugation are given respectively by $X\rightarrow \sigma_{x}^{T}
X^{T}\sigma_{x}$ and $X\rightarrow J_{2}^{T}X^{T}J_{2}$, and then
applying {\bf IC} to this data yields the following:
\begin{itemize}
\item A basis of $1$-vectors $\{ \sigma_{z}\otimes \sigma_{x},
\sigma_{x}\otimes I_{2}, \sigma_{z}\otimes i\sigma_{y},
i\sigma_{y}\otimes I_{2}\}$ for $Cl(2,2)$
\item Reversion $Cl(2,2)$, with respect to this basis, is
$X\rightarrow K_{4}^{T} \left (\begin{array}{cc}
J_{2} & 0\\
0 & J_{2}
\end{array}\right )^{T}X^{T} \left (\begin{array}{cc}
J_{2} & 0\\
0 & J_{2}
\end{array}\right )K_{4}$, i.e., it is $X\rightarrow 
M_{1\otimes k}^{T}X^{T}M_{1\otimes k}$.
\item Clifford conjugation on $Cl(2,2)$, with respect to this basis,
is given by $X\rightarrow J_{4}^{T}
\left (\begin{array}{cc}
\sigma_{x} & 0\\
0 & \sigma_{x}
\end{array}
\right )^{T}X^{T}
\left (\begin{array}{cc}
\sigma_{x} & 0\\
0 & \sigma_{x}
\end{array}
\right )J_{4}
= 
M_{-k\otimes 1}^{T}X^{T}M_{-k\otimes 1}$
\item Hence the grade automorphism
on $Cl(2,2)$, with respect to this basis of $1$-vectors,
is $X\rightarrow 
M_{k\otimes k}^{T}XM_{k\otimes k}$
\end{itemize}

Now $M_{k\otimes k}$ is precisely $\left (\begin{array}{cc}
\sigma_{z} & 0\\
0 & -\sigma_{z}
\end{array}
\right ) = \sigma_{z}\otimes \sigma_{z}$. So $X\in Cl(2,2) = 
M(4, \mathbf{R})$ is
even its blocks $X_{ij}, i,j=1,2$ satisfy
\[
\left (\begin{array}{cc}
X_{11}\sigma_{z} & -X_{12}\sigma_{z}\\
X_{21}\sigma_{z} & -X_{22}\sigma_{z}
\end{array}
\right ) =
\left (\begin{array}{cc}
\sigma_{z}X_{11} & \sigma_{z}X_{12}\\
-\sigma_{z}X_{21} & -\sigma_{z}X_{22}
\end{array}
\right )
\]
So $X_{11}, X_{22}$ commute with $\sigma_{z}$ and $X_{12}, X_{21}$ anticommute
with $\sigma_{z}$. Equivalently, $X_{11}, X_{22}$ are boh diagonal and
$X_{12}, X_{22}$ are both anti-diagonal.

Thus, an even vector $X$ in $Cl(2,2) = M(4, \mathbf{R})$ is given by
\[
X = \left (\begin{array}{cccc}
a & 0 & 0 & b\\
0 & d & c & 0\\
0 & \beta & \alpha & 0\\
\gamma & 0 & 0 & \delta
\end{array}
\right )
\]

Such an even $X$ is in ${\mbox Spin}^{+}(2,2)$ iff it satisfies
$X^{cc}X = I_{4}$. Equivalently
\[
\left (\begin{array}{cc}
0 & \sigma_{x}\\
-\sigma_{x} & 0
\end{array}
\right ) \left (\begin{array}{cc}
X_{11}^{T} & X_{21}^{T}\\
X_{12}^{T} & X_{22}^{T} 
\end{array}
\right ) \left (\begin{array}{cc}
0 & -\sigma_{x}\\
\sigma_{x} & 0
\end{array}
\right ) \left (\begin{array}{cc}
X_{11} & X_{12}\\
X_{21} & X_{22} 
\end{array}
\right ) = I_{4}
\]
This, in turn is equivalent to the following 4 conditions:
\begin{itemize}
\item $(1,1)$ block $ = 
I_{2}$. This is the same
as requiring $\left (\begin{array}{cc}
a\delta - b\gamma & 0\\
0 & \alpha \delta -c\beta 
\end{array}
\right ) = I_{2}$

\item $(1,2)$ block $=0_{2}$. But the $(1,2)$ block is
equals $\left (\begin{array}{cc}
0 & b\delta - b\delta \\
c\alpha - c\alpha & 0
\end{array}
\right )$. So this is no constraint.
\item  Similarly, $(2,1)$ block $= 0_{2}$
is no constraint either.

\item $(2,2)$ block $ = I_{2}$. The $(2,2)$ block is
$\sigma_{x}X_{11}^{T}\sigma_{x}X_{22} - \sigma_{x}X_{21}^{T}\sigma_{x}X_{12}$,  which equals $\left (\begin{array}{cc}
\alpha d - \beta c & 0\\
 0 & a\delta - b\gamma 
\end{array}
\right )$. But this is precisely the first constraint again.
\end{itemize}
Hence we can conclude that $X =  \left (\begin{array}{cccc}
a & 0 & 0 & b\\
0 & d & c & 0\\
0 & \beta & \alpha & 0\\
\gamma & 0 & 0 & \delta
\end{array}
\right )
$ is in ${\mbox Spin}^{+}(2,2)$ iff the matrices $\left (\begin{array}{cc}
a & b\\
\gamma & \delta
\end{array}
\right )$ and $\left (\begin{array}{cc}
d & c\\
\beta & \alpha 
\end{array}
\right )$ are both in $SL(2, R)$. Thus ${\mbox Spin}^{+}(2,2)$
is isomorphic to $SL(2, \mathbf{R})\times SL(2, \mathbf{R})$ 
embedded {\it concentrically}
in $M(4, \mathbf{R})$.          

\subsection{${\mbox Spin}^{+}(3,3)$}
Applying {\bf IC} to the basis of $1$ vectors for $Cl (2,2)$
in the previous subsection
leads to the following
initial basis of $1$-vectors for $Cl(3,3)$:
\begin{equation}
\label{IBasis33}
\{ \sigma_{z}\otimes \sigma_{z}\otimes \sigma_{x}, 
\sigma_{z}\otimes \sigma_{x}\otimes I_{2},
\sigma_{x}\otimes I_{2}\otimes I_{2},
\sigma_{z}\otimes \sigma_{z}\otimes i\sigma_{y},
\sigma_{z}\otimes i\sigma_{y}\otimes I_{2},
i\sigma_{y}\otimes I_{2}\otimes I_{2}\}
\end{equation}

A calculation plus the observation that 
$\left (\begin{array}{cc} 
0 & M_{1\otimes k}\\
-M_{1\otimes k} & 0
\end{array}
\right ) = i\sigma_{y}\otimes \sigma_{x}\otimes i\sigma_{y}$,
shows that with respect to this basis Clifford conjugation on $Cl(3,3)$ is
\[
X\rightarrow (i\sigma_{y}\otimes \sigma_{x}\otimes i\sigma_{y})^{T}
X^{T} (i\sigma_{y}\otimes \sigma_{x}\otimes i\sigma_{y})
\]

Similarly, using the observation that
 $M_{k\otimes 1} = i\sigma_{y}\otimes \sigma_{x}$,
one finds that reversion, with respect to this basis, is given by
\[
X\rightarrow (\sigma_{x}\otimes i\sigma_{y}\otimes \sigma_{x})^{T}
X^{T} (\sigma_{x}\otimes i\sigma_{y}\otimes \sigma_{x})
\]

Hence, using basic properties of the Kronecker product,
it is seen that the grade involution, with respect to this basis, is
\[
X\rightarrow  (\sigma_{z}\otimes \sigma_{z}\otimes \sigma_{z})^{T}
X(\sigma_{z}\otimes \sigma_{z}\otimes \sigma_{z}) 
= \sigma_{z}\otimes \sigma_{z}\otimes \sigma_{z}X
  \sigma_{z}\otimes \sigma_{z}\otimes \sigma_{z}
\]

Unlike the case of ${\mbox Spin}^{+}(2,2)$ the block structure
of the even vectors in $Cl(3,3)$ with respect to this choice of
the grade morphism is no longer very informative. So we look for
an explicit conjugation to a more palatable version of the grade morphism.
Now $\sigma_{z}\otimes \sigma_{z}\otimes \sigma_{z}$ is diagonal with
four $1$s and four $-1$s on the diagonal. Hence, by inspection 
it is explicitly
permutation symmetric to ${\mbox diag} (I_{4}, -I_{4})$.
Specifically, if 
\[
P = [e_{1}\mid e_{4} \mid e_{6}\mid e_{7}\mid e_{2}\mid e_{3}\mid e_{5}
\mid e_{8}]
\]
Then $P^{T}(\sigma_{z}\otimes\sigma_{z}\otimes\sigma_{z})P
= {\mbox diag}(I_{4}, -I_{4})$.

We would like the grade involution to be given $X\rightarrow
{\mbox diag}(I_{4}, -I_{4})X{\mbox diag}(I_{4}, -I_{4})$, so that an even
vector is then precisely the set of all \underline{block-diagonal}
$8\times 8$ matrices [cf., Remark (\ref{CommutewithDiagonal})].
.
To that end we change basis of $1$ vectors according to
\[
X_{i} = P^{T}Y_{i}P, 
i =1, \ldots, 6 
\]
where the $Y_{i}$ form the basis in Equation (\ref{IBasis33}).

Then a calculation provides the basis $\{X_{i}\}, i=1, \ldots , 6$ via
\begin{equation}
\label{IIBasis33}
\{\sigma_{x}\otimes \sigma_{z}\otimes I_{2}, \sigma_{x}\otimes\sigma_{z}
\otimes\sigma_{x}, \sigma_{x}\otimes\sigma_{x}\otimes I_{2},
i\sigma_{y}\otimes I_{2}\otimes I_{2}, \sigma_{x}\otimes \sigma_{z}\otimes
i\sigma_{y}, \sigma_{x}\otimes i\sigma_{y}\otimes I_{2}\}
\end{equation}

Then Clifford conjugation, with respect to the basis
$X_{i}$, is given by $X\rightarrow C_{3,3}^{T}X^{T}C_{3,3}$,
where we have
\[
C_{3,3} = P^{T}(i\sigma_{x}\otimes\sigma_{x}\otimes i\sigma_{y})P
\]

The matrix $C_{3,3}$ is given a neat form by availing of the observation
that  
$i\sigma_{x}\otimes\sigma_{x}\otimes i\sigma_{y}
= [e_{8}\mid -e_{7}\mid e_{6}\mid -e_{5}\mid -e_{4}\mid e_{3}
\mid -e_{2}\mid e_{1}]$.

Hence,  
\begin{equation}
\label{Cliff33}
C_{3,3} = \left (\begin{array}{cc}
0_{4} & M_{1\otimes k}\\
-M_{1\otimes k} & 0_{4}
\end{array}
\right )
\end{equation}

So ${\mbox Spin}^{+}(3,3)$ is a subgroup of block-diagonal $8\times 8$
matrices $\left (\begin{array}{cc}
A & 0_{4}\\
0_{4} & D
\end{array}
\right )$ satisfying two additional conditions:
\begin{itemize}
\item $C_{3,3}^{T}\left (\begin{array}{cc}
A & 0_{4}\\
0_{4} & D
\end{array}
\right )^{T} C_{3,3} \left (\begin{array}{cc}
A & 0_{4}\\
0_{4} & D
\end{array}
\right ) = I_{8}$.
\item $\left (\begin{array}{cc}
A & 0_{4}\\
0_{4} & D
\end{array}
\right ) X \left (\begin{array}{cc}
A & 0_{4}\\
0_{4} & D
\end{array}
\right )^{CC}$ is a $1$-vector for all $1$-vectors $X$.
\end{itemize}

The first of these conditions is equivalent to the condition that 
$
-M_{1\otimes k}A^{T}M_{1\otimes k}D = I_{4}$.
So $A\in GL(4, \mathbf{R})$ can be taken 
to be arbitrary and then $D$ is prescribed
as $D = -M_{1\otimes k}A^{-T}M_{1\otimes k}$.    
  
\begin{remark}
\label{linearized}
{\rm It is useful to write down the linearization of the conditions
in the last paragraph, because that describes the Lie algebra of
${\mbox Spin}^{+}(3,3)$. The linearization of these conditions describes
a block-diaogonal $\left (\begin{array}{cc}
R & 0\\
0 & S 
\end{array}
\right )$, with $R, S\in M(4, R)$ satisfying 
\[
\left (\begin{array}{cc}
R^{T} & 0_{4}\\
0_{4} & S^{T}
\end{array}
\right ) \left (\begin{array}{cc}
0_{4} & M_{1\otimes k}\\
-M_{1\otimes k} & 0_{4}
\end{array}
\right ) = - \left (\begin{array}{cc}
0_{4} & M_{1\otimes k}\\
-M_{1\otimes k} & 0_{4}
\end{array}
\right )\left (\begin{array}{cc}
R & 0_{4}\\
0_{4} & S
\end{array}
\right )
\]
This is equivalent to $R^{T}M_{1\otimes k} = -M_{1\otimes k}S$ and 
$-S^{T}M_{1\otimes k} = M_{1\otimes k}R$.
Once again these two conditions are equivalent to one another and
thus to $S = M_{1\otimes k}R^{T}M$. Thus ${\mbox spin}^{+}(3,3)$
is the collection of block-diagonal matrices $Z =
\left (\begin{array}{cc}
R & 0\\
0 & M_{1\otimes k}R^{T}M_{1\otimes k}
\end{array}
\right )$ satisfying the additional condition that $ZX - XZ$ is a $1$
-vector for all $1$ vectors.}
\end{remark} 

Let us use this to show that ${\mbox Spin}^{+}(3,3)$ is indeed $SL(4, R)$.
For this, using Remark (\ref{DiscernTraceFree}),
it is easier to work with the linearized version of the
condition $\left (\begin{array}{cc}
A & 0\\
0 & D
\end{array}
\right ) X \left (\begin{array}{cc}
A & 0\\
0 & D
\end{array}
\right )^{CC}$ is a $1$-vector for all $1$-vectors, and show that it is 
violated by the prototypical element of $M(4, \mathbf{R})$ which is not in
$sl(4, \mathbf{R})$ - viz., $I_{4}$ or rather its embedding as an element
of $M(8, \mathbf{R})$ satisfying the conditions in Remark (\ref{linearized}).
Since $M_{1\otimes k}I_{4}M_{1\otimes k} = -I_{4}$ we have to show
that $Z = \left (\begin{array}{cc}
I_{4} & 0_{4}\\
0_{4} & -I_{4}
\end{array}
\right ) = \sigma_{z}\otimes I_{2}\otimes I_{2}$ 
violates the condition $ZX -XZ$ is a $1$-vector for all
$1$-vectors $X$. This is easily verified by using elementary properties
of the Kronecker product - especially {\bf KP3}. 

Hence we conclude that ${\mbox spin}^{+}(3, 3)  
= \{ \left (\begin{array}{cc}
R & 0\\
0 & M_{1\otimes k}R^{T}M_{1\otimes k}
\end{array}
\right ), R\in sl(4, \mathbf{R})\}$. Hence, ${\mbox Spin}^{+}(3,3)$
is a group of $8\times 8$ real matrices of the form
$\{ \left (\begin{array}{cc}
A & 0\\
0 & -M_{1\otimes k}A^{-T}M_{1\otimes k}
\end{array}
\right ), A\in SL(4, \mathbf{R}) \}$. This group is evidently isomorphic
to $SL(4, \mathbf{R})$.

\subsection{${\mbox Spin}^{+}(4,1)$}

Consider the following basis of $1$-vectors for $Cl (3, 0) = M(2, \mathbf{C})$: 
\[
\{\sigma_{x}, \sigma_{y}, \sigma_{z}\}
\]

Reversion with respect to this basis of $1$-vectots is given by
$X\rightarrow X^{*}$ and Clifford conjugation is $X\rightarrow 
J_{2}^{T} X^{T}J_{2}$. Therefore {\bf IC} applied to this data yields
the following:
\begin{itemize}
\item A basis of $1$-vectors for $Cl(4, 1) = M(4, \mathbf{C})$:
\[
\{Y_{1} = \sigma_{z}\otimes \sigma_{x}, Y_{2} = \sigma_{z}\otimes \sigma_{y},
Y_{3} = \sigma_{z}\otimes \sigma_{z}, Y_{5} = K_{4} =
\sigma_{x}\otimes I_{2}, Y_{5} = J_{4} = i\sigma_{y}\otimes I_{2}\}
\]
\item Clifford conjugation with respect to this basis is
$X\rightarrow J_{4}^{T}
X^{*}J_{4}$
\item Reversion with respect to this basis is
$X\rightarrow  M_{ 1 \otimes k}^{T}
X^{T} M_{1\otimes k}$.
\item Hence the grade involution, with respect to this basis of $1$-vectors,
is given by $X\rightarrow (J_{4}M_{1\otimes k})^{T}\bar{X}  
(J_{4}M_{1\otimes k})$. Since $J_{4} = M_{1\otimes j}$, the grade involution
is $X\rightarrow M_{1\otimes i}^{T}\bar{X}  
M_{1\otimes i}$.
\end{itemize}

In view of $Cl(4, 1)$ being the algebra of an even sized (i.e., 4) complex
matrices, it is prudent to seek a basis of $1$-vectors with respect
to which the grade involution is given by $X\rightarrow
J_{4}^{T}\bar{X}J_{4}$. Now $J_{4} = M_{1\otimes j}
= M_{1\otimes q}^{T}M_{1\otimes i}M_{1\otimes q}$,
for a unit quaternion $q$ satisfying $\bar{q}iq = j$. For instance,
pick $q = \frac{1}{\sqrt{2}} (1-k)$.
Accordingly we change the basis of $1$-vectors for $Cl(4, 1)$ to
$X_{i} = M_{1\otimes q}^{T}Y_{i}M_{1\otimes q}$.

Then with respect to this basis of $1$-vectors:
\begin{itemize}
\item Grade involution is
$X\rightarrow M_{1\otimes j}^{T}\bar{X}M_{1\otimes j}$. 
\item Clifford conjugation is given by
$X\rightarrow (M_{1\otimes q}^{T}J_{4}M_{1\otimes q})^{T}
X^{*}(M_{1\otimes q}^{T}J_{4}M_{1\otimes q})$.
But $M_{1\otimes q}^{T}J_{4}M_{1\otimes q} = M_{1\otimes -i}$ as can
be seen quickly from quaternionic algebra.  
So Clifford conjugation with respect to this basis is  $X\rightarrow
M_{1\otimes i}^{T}X^{*}M_{1\otimes i}$
\end{itemize}

Thus, even vectors precisely those matrices in $M(4, \mathbf{C})$ which are
in the image of $\Theta_{H}$ and thus we have

\begin{proposition}
{\rm ${\mbox Spin}^{+}(4, 1) = \{ X\in M(4, \mathbf{C}): X\in Im (\Theta_{H});
M_{1\otimes i}^{T}X^{*}M_{1\otimes i}X = I_{4}\}$.}
\end{proposition}     

\subsection{${\mbox Spin}^{+}(1, 5)$}

Let us start with 
the following basis of $1$-vectors for $Cl (0, 4) = M(2, \mathbf{H})$:
\begin{eqnarray*}
Z_{1} & = & \left (\begin{array}{cc}
0 & i\\
i & 0
\end{array}
\right ) \\ \nonumber
Z_{2} & = & \left (\begin{array}{cc}
0 & j\\
j & 0
\end{array}
\right ) \\ \nonumber
Z_{3} & = & \left (\begin{array}{cc}
0 & k\\
i & 0
\end{array}
\right ) \\ \nonumber
Z_{4} & = & \left (\begin{array}{cc}
0 & 1\\
-1 & 0
\end{array}
\right ) 
\end{eqnarray*}

With respect to this basis of 
$1$- vectors, $Z_{i}$, some calculations reveal
that : i) Clifford Conjugation is given
by $X\rightarrow X^{*}$; ii) Reversion is given by $X\rightarrow
\sigma_{z}X^{*}\sigma_{z}$; and hence iii) the grade involution
is expressible as $X\rightarrow \sigma_{z}X\sigma_{z}$. 

Applying {\bf IC} to this basis yields a basis $\{Y_{i}\}, i=1, \ldots , 6$
of $1$-vectors
for $Cl(1, 5)$, given by

\[
\sigma_{x}\otimes I_{2}, 
\sigma_{z}\otimes \left (\begin{array}{cc}
0 & i\\
i & 0
\end{array}
\right ), 
\sigma_{z}\otimes \left (\begin{array}{cc}
0 & j\\
j & 0
\end{array}
\right ),
\sigma_{z} \otimes \left (\begin{array}{cc}
0 & k\\
i & 0
\end{array}
\right ),
\sigma_{z}\otimes \left (\begin{array}{cc}
0 & 1\\
-1 & 0
\end{array}
\right ),
i\sigma_{y}\otimes I_{2} 
\]
                      
{\bf IC} then also shows that, with respect to the basis of $1$-vectors
$Y_{i}$:
\begin{itemize}
\item Clifford conjugation is given by $X\rightarrow
J_{4}^{*} \left (\begin{array}{cc}
\sigma_{z}^{*} & 0\\
0 & \sigma_{z}^{*}
\end{array}
\right ) X^{*} \left (\begin{array}{cc}
\sigma_{z} & 0\\
0 & \sigma_{z}
\end{array}
\right )J_{4}
= \left (\begin{array}{cc}
0 & \sigma_{z}\\
-\sigma_{z} & 0
\end{array}
\right )^{*}X^{*} \left ( \begin{array}{cc}
0 & \sigma_{z}\\
-\sigma_{z} & 0
\end{array}
\right )$
\item Reversion is given by
$X\rightarrow K_{4}^{*}X^{*}K_{4}$
\item Hence the grade involution is
$ X \rightarrow [\left (\begin{array}{cc}
0 & \sigma_{z}\\
-\sigma_{z} &  0
\end{array}
\right )K_{4}]^{*}  
X \left (\begin{array}{cc}
0 & \sigma_{z}\\
-\sigma_{z} &  0
\end{array}
\right )K_{4}=
\left (\begin{array}{cc}
\sigma_{z} & 0\\
0 & -\sigma_{z}
\end{array}
\right )X \left (\begin{array}{cc}
\sigma_{z} & 0\\
0 & -\sigma_{z}
\end{array}
\right )$
\end{itemize}

Once again by Remark (\ref{CommutewithDiagonal}),
it would be preferable to
have the grade involution described by
$X\rightarrow {\mbox diag}( I_{2}, -I_{2})X
{\mbox diag}( I_{2}, -I_{2})$, so that even vectors become block-diagonal
matrices. This is achieved by mere inspection, since
$P^{T} \left (\begin{array}{cc}
\sigma_{z} & 0\\
0 & -\sigma_{z}
\end{array}
\right )P = {\mbox diag}( I_{2}, -I_{2})$,
where $P$ is permutation matrix
\[
P = [e_{1}\mid e_{4}\mid e_{2}\mid e_{3}
]
\]

Accordingly we change the basis of $1$-vectors to $X_{i}$, with
$X_{i} = P^{T}Y_{i}P$.
We then find, after some block matrix manipulations that
\begin{itemize}
\item $X_{1} = P^{T}Y_{1}P =
\sigma_{x}\otimes \sigma_{x}$ 
\item $X_{2} = P^{T}Y_{2}P 
=\sigma_{x}\otimes i\sigma_{z}$
\item $X_{3} = P^{T}Y_{3}P 
= i\sigma_{y}\otimes I_{2} = J_{4}$
\item $Y_{4} = P^{T}X_{4}P = \sigma_{x}\otimes j\sigma_{z}$. Here the $j$
is the quaternion unit $j$.
\item $Y_{5}  = P^{T}X_{5}P = \sigma_{x}\otimes k\sigma_{z}$.
Similarly, the $k$ is the quaternion unit $k$.
\item $X_{6} = P^{T}Y_{6}P =
\sigma_{x}\otimes i\sigma_{y}$  
\end{itemize} 

Clifford conjugation, with respect to this basis of $1$-vectors, is given
by $X\rightarrow C_{1,5}^{*}X^{*}C_{1,5}$, where $C_{1,5}
= P^{T}\left (\begin{array}{cc}
0 & \sigma_{z}\\
-\sigma_{z} & 0
\end{array}
\right )P = \left (\begin{array}{cccc}
0 & 0 & 0 & 1\\
0 & 0 & 1& 0\\
0 & -1 & 0 & 0\\
-1 & 0 & 0 & 0
\end{array}
\right ) = i\sigma_{y}\otimes \sigma_{x}$.

Thus ${\mbox Spin}^{+}(1, 5)$ consists of those $X\in M(4, \mathbf{H})$
which satisfy : i) $X$ is block-diagonal ${\mbox diag}(X_{1}, X_{2})$,
with each $X_{i}\in M(2, \mathbf{H})$, ii) $X^{cc}X = I_{4}$, and iii)
$X^{cc}ZX$ is a $1$-vector for all $1$-vectors $Z$.

Focussing on the condition, $X^{cc}X = I_{4}$, 
it is found that it is equivalent to
$\sigma_{x}X_{2}^{*}\sigma_{x}X_{1} = I_{2}$ and $\sigma_{x}X_{1}^{*}
\sigma_{x}X_{2} = I_{2}$. But the first of these two conditions is equivalent
to the second, as follows by $*$-conjugating the first
equation and then pre and post-multiplying the result by $\sigma_{x}$.
Finally, the first of these conditions is equivalent to$X_{1}$ being
invertible and $X_{2}^{-1} = \sigma_{x}X_{1}^{*}\sigma_{x}$;
and iii) $X^{cc}ZX$ is a $1$-vector for all $1$-vectors $Z$.

Thus, ${\mbox Spin}^{+}(1,5)$ is isomorphic to the collection of matrices
in $M(4, \mathbf{H})$ satisfying:
\begin{enumerate} 
\item $X$ is block-diagonal with diagonal blocks
 $X_{1}, X_{2}$ both
invertible in $M(2, \mathbf{H})$, and 
$X_{2}^{-1} = \sigma_{x}X_{1}^{*}\sigma_{x}$, and 
\item $X^{cc}ZX$ is a $1$-vector for all $1$-vectors $Z$.
\end{enumerate}

To identify the restriction imposed by the last condition it is
easier to work with the Lie algebra ${\mbox spin}^{+}(1, 5)$.
This should be some subalgebra of block-diagonal matrices
${\mbox diag}(A_{1}, A_{2})$, with $A_{i}\in M(2, H)$ satisfying
$A_{2} = \sigma_{x}A_{1}^{*}\sigma_{x}$. To specify this subalgebra, we
use by way of variation the second alternative in 
Remark (\ref{DiscernTraceFree}), i.e.,
the space of bivectors (with respect to
the basis of $1$-vectors $X_{i}$) is explicitly computed, since this equals 
${\mbox spin}^{+}(1, 5)$.

This computation yields the following as the basis of the space of
bivectors:
\begin{enumerate}
\item $X_{1}X_{2} = I_{2}\otimes \left (\begin{array}{cc}
0 & -i\\
i & 0
\end{array}
\right )$;
$X_{1}X_{3} = I_{2}\otimes \left (\begin{array}{cc}
0 & -j\\
j & 0
\end{array}
\right )$;
$X_{1}X_{4} = I_{2}\otimes \left (\begin{array}{cc}
0 & -k\\
k & 0
\end{array}
\right )$;
$X_{1}X_{5} = -\sigma_{x}\otimes \sigma_{x}$;
$X_{1}X_{6} = I_{2}\otimes -\sigma_{z}$.
\item $X_{2}X_{3} = I_{2}\otimes kI_{2}$;
$X_{2}X_{4} = -I_{2}\otimes jI_{2}$;
$X_{2}X_{5} = -\sigma_{z}\otimes i\sigma_{z}$;
$X_{2}X_{6} = I_{2}\otimes i\sigma_{x}$.
\item $X_{3}X_{4} = I_{2}\otimes iI_{2}$;
$X_{3}X_{5} = -\sigma_{z}\otimes j\sigma_{z}$;
$X_{3}X_{6} = I_{2}\otimes j\sigma_{x}$.
\item $X_{4}X_{5}= -\sigma_{z}\otimes k\sigma_{z}$;
$X_{4}X_{6} = I_{2}\otimes k\sigma_{x}$.
\item $X_{5}X_{6} = \sigma_{z}\otimes i\sigma_{y}$.
\end{enumerate}

Thus,
\begin{itemize}
\item The typical element in the span of $\{X_{1}X_{2}, X_{1}X_{3},
X_{1}X_{4}\}$ is given by $I_{2}\otimes \left (\begin{array}{cc}
0 & p\\
\bar{p} & 0
\end{array}
 \right )$, with $p\in \mathbf{P}$.
\item The typical element in the span of $\{X_{2}X_{5}, X_{3}X_{5},
X_{4}X_{5}\}$ is given by $-\sigma_{z}\otimes \left (\begin{array}{cc}
q & 0\\
0 & \bar{q} 
\end{array}
\right )$, with $q\in \mathbf{P}$. 
 
\item The typical element in the span of $\{X_{2}X_{6}, X_{3}X_{6},
X_{4}X_{6}\}$ is given by $I_{2}\otimes \left (\begin{array}{cc}
0 & p\\
p & 0
\end{array}
\right )$, with $s\in \mathbf{P}$.
\item The typical element in the span of $\{X_{1}X_{6}, X_{2}X_{3},
X_{2}X_{4}, X_{3}X_{4}\}$ is given by $I_{2}\otimes \left (\begin{array}{cc}
\alpha  & 0\\
0 & -\bar{\alpha}
\end{array}
\right )$, with $\alpha \in \mathbf{H}$.

\item The typical element in the span of $\{X_{1}X_{5}, X_{5}X_{6}\}$,
is given by $\sigma_{z}\otimes \left (\begin{array}{cc}
0 & r_{1}\\
r_{2} & 0
\end{array}
 \right )$, with $r_{1}, r_{2}\in R$.
\end{itemize}

Thus, a typical bivector is a block-diagonal matrix, with each block
in $M(2, \mathbf{H})$. 
Furthermore, the NW block of a bivector, which also determines
the SE block,
is of the form:
\[
\left (\begin{array}{cc}
\alpha - q & p + s + r_{1}\\
\bar{p} + s + r_{2} & -\bar{q} - \bar{\alpha}
\end{array}
\right )
\]

This is clearly of the form
\[
\left (\begin{array}{cc}
q_{11} & q_{12}\\
q_{21} & q_{22} 
\end{array}
\right )
\]

with $q_{ij}\in \mathbf{H}$ arbitrary, 
except for ${\mbox Re} (q_{11} + q_{22})
= 0$. Thus, in view of Remark (\ref{SLNH}), the NW block, $X_{11}$,
can be chosen to be any element of $sl (2, \mathbf{H})$, 
while the SE block, $X_{22}$
is prescribed by $X_{22} = -\sigma_{x}X_{11}^{*}\sigma_{x}$.
This confirms that ${\mbox spin}^{+}(1,5)$ is indeed isomorphic
to $sl(2, \mathbf{H})$ whence
${\mbox Spin}^{+} (1, 5)$ is isomorphic to $SL(2, \mathbf{H})$.

\section{Conclusions}

In this work we have developed what we consider to be a straightforward
approach to the indefinite spin groups ${\mbox Spin}^{+}(p, q)$ for
$p+q\leq 6$. By working in the matrix algebra that $Cl(p, q)$ is isomorphic
to, as opposed to the matrix algebra that its even subalalgebra 
is isomorphic to,
one can now address applications requiring the spin group to live in the
same matrix algebra as the one-vectors (thereby eliminating 
ad hoc constructions). Furthermore the method has a flavour which is
closer to the defining conditions for the spin groups making it, 
in our opinion, didactically more palatable. As byproducts we find
novel representations for some of the classical groups, such as the
$\Lambda$ embedding of $\breve{Sp}(4, \mathbf{R})$ in $M(8, \mathbf{R})$
and the concentric embedding of $SL(2, \mathbf{R})\times SL(2, \mathbf{R})$
in $SL(4, \mathbf{R})$, and more natural interpretations of some of the
members of the $\mathbf{H}\otimes \mathbf{H}$ basis of $M(4, \mathbf{R})$.
Some obvious questions arise out of this work:
\begin{itemize}
\item Though ${\mbox Spin}^{+}(q, p)$ is isomorphic to ${\mbox Spin}^{+}
(p,q)$, the same is not true for $Cl(q, p)$ and $Cl(p, q)$. Thus, it would
be a profitable exercise to apply this approach to study ${\mbox Spin}^{+}
(q, p)$ corresponding to the pairs $(p, q)$ studied in this work.
It is reasonable to expect this to lead to more novel representations for
some of the classical groups.

\item A natural question is how much further
this method extends to higher dimensions.
The form of the iteration {\bf IC} in conjunction with basic block matrix
algebra renders the  extension of the
benefits of the quaternion tensor product 
a distinct possibility in higher dimensions
as well (indeed, even in this work
most of the $Cl(p,q)$s were  isomorphic
to algebras of matrices of size already larger than $4\times 4$).
Thus finding bases of $1$-vectors, matrix forms of Clifford conjugation
etc., ought to be possible.
The real ``bottleneck" is the analysis of the third condition in
the definition of the spin groups [See IV) of Definition (\ref{basics})].
The methods outlined in Remark (\ref{DiscernTraceFree}) would have to
be substantially extended to this end.

\item It would be worthwhile to complete the analysis started in this paper
to compute exactly and explicitly
the exponentials of matrices in Lie algebras other than 
${\mbox spin}(3,2, \mathbf{R})$. Of course, some of the other 
$so (p, q, \mathbf{R})$s, such $so (2, \mathbf{R})$ etc., are even simpler,
since now one has to exponentiate only $2\times 2$ matrices.
But the explicit computation
of the minimal polynomials of matrices in $su(2,2)$, $SL(4,
\mathbf{R})$ etc., - a worthwhile task with applications going 
beyond exponentiation - is needed for the other
$so (p, q, \mathbf{R})$s and will require significant effort. 
\end{itemize}

\section{Appendix - Spin Groups and Matrix Exponentiation}
Computing the exponential of a matrix is central to much 
of applied mathematics. In general, this is quite arduous, \cite%
{dubious}. However, for matrices $X \in so (p, q, \mathbf{R})$,
the theory of Clifford
Algebras and spin groups enables the reduction of finding $e^{X}$, with $%
X\in so (p, q, \mathbf{R})$, 
to the computation of $e^{Y}$, where $Y$ is the associated
element in the Lie algebra ${\mbox spin}^{+}(p,q)$, 
via the method detailed below. Typically $Y$
is smaller in size than $X$. 
In particular, the minimal
polynomial of $Y$ is \underline{typically of lower degree} than that of $X$
and often is also more structured.
The method is described in the following algorithm, which is an adaptation
of a similar algorithm in \cite{jgspi}: 

\begin{algorithm}

\begin{description}
\item[Step 1] Identify a collection of matrices which serve as a basis of $1$
vectors for the Clifford Algebra $Cl (p, q) $.

\item[Step 2] Identify the explicit form of Clifford conjugation ($\phi ^{cc}
$) and the grade (or so-called main) automorphism on $Cl (p, q)$
with respect to this collection of matrices.
Equivalently identify the explicit form of Clifford conjugation and
reversion ($\phi ^{rev}$) with respect to this collection of matrices.

\item[Step 3] Steps 1 and 2 help in identifying both the spin group 
${\mbox Spin}^{+}(p, q)$ and its Lie algebra ${\mbox spin}^{+}
(p, q)$ as
sets of matrices, within the same matrix algebra, that the
matrices in Step 1 live in. Hence, the double
covering $\Phi _{p, q}: {\mbox Spin}^{+}(p, q)
\rightarrow SO^{+}(p, q, \mathbf{R})$
the matrix, with respect to the basis
of $1$-vectors in Step 1, of the linear map $H\rightarrow ZH\phi ^{cc}(Z)$,
with $H$ a matrix in the collection of $1$-vectors in \emph{Step 1} and 
$Z\in {\mbox Spin}^{+}(p, q)$. This enables one to express $\Phi _{p, q}(Z)$
as a matrix in $SO^{+}(p, q, \mathbf{R})$.

\item[Step 4] Linearize $\Phi _{p, q}$ to obtain 
the Lie algebra isomorphism $\Psi_{p, q}
:{\mbox spin}^{+} (p, q) \rightarrow so (p, q, \mathbf{R})$. 
This reads as $W\rightarrow YW-WY$, with $W$ once again a $1$%
-vector and $Y\in {\mbox spin}^{+}(p, q)$. Once again this leads to a
matrix in $so (p, q, \mathbf{R})$, 
which is $\Psi _{p, q}(Y)$.

\item[Step 5] Given $X\in so(p, q, \mathbf{R})$,\text{ }%
find $\Psi _{p, q}^{-1}(X)= Y\in {\mbox spin}^{+}(p, q)$.

\item[Step 6] Compute the matrix $e^{Y}$ and use Step 3 to find the matrix $%
\Phi _{p, q}(e^{Y})$. This matrix is $e^{X}$.
\end{description}
\end{algorithm}

With the contents of this work, this algorithm will be executable
for $X\in so (p,q, \mathbf{R})$, for those pairs $(p, q)$ considered
in this work, as long as there are closed form formulae for finding
the $e^{Y}$ of Step 6 above. For each such pair, $(p, q)$, considered
in this work, the corresponding $Y$ is atmost $4\times 4$. So the minimal
polynomials of such $Y$ is atmost a  quartic. For some pairs (such as 
$(p, q)= (3, 2)$, for instance) the list of possible minimal polynomials
has even greater structure, \cite{minpolyi},
and thus the attendant problem of exponentiation becomes even simpler.
Obtaining the minimal polynomials of such $Y$'s is left to future work.

\end{document}